\documentclass[fleqn,usenatbib]{mnras}

\usepackage{newtxtext,newtxmath}

\usepackage[T1]{fontenc}
\usepackage{ae,aecompl}

\usepackage{graphicx}	
\usepackage{amsmath}	
\usepackage{amssymb}	

\title[Study of Li-rich giants with the GALAH survey]{Study of Lithium Rich Giants with the GALAH Spectroscopic Survey}

\author[Deepak and Reddy, B. E.]{
Deepak$^{1}$\thanks{E-mail: Deepak@iiap.res.in}
and Bacham E. Reddy$^{1}$
\\
$^{1}$Indian Institute of Astrophysics, Bangalore 560034, India
}

\date{Accepted January 10, 2019. Received January 10, 2019; in original form December 12, 2018}

\pubyear{2018}

\begin{document}
\label{firstpage}
\pagerange{\pageref{firstpage}--\pageref{lastpage}}
\maketitle

\begin{abstract}
In this article, we speculate on the possible mechanisms for Li enhancement origin in RGB stars based on a large data set of around 340,299 stars collected from the GALAH survey combined with the Gaia astrometry. Data has 51,982 low mass (M$\leq$ 2M$_{\odot}$) RGB stars with reliable atmospheric parameters. The data set shows a well populated RGB with well-defined luminosity bump and red clump with significant number of stars at each of these two key phases. We found 335 new Li-rich RGB stars with Li abundance, A(Li) $\geq$ 1.80 $\pm$ 0.14 dex, of which 20 are super Li-rich with A(Li) $\geq$ 3.20~dex. Most of them appear to be in the red clump region which, when combined with stellar evolutionary timescales on RGB, indicates that the Li enhancement origin may lie at RGB tip during He-flash rather than by external source of merging of sub-stellar objects or during luminosity bump evolution. Kinematic properties of sample stars suggest that Li-rich giants are relatively more prevalent among giants of thin disk compared to thick disk and halo.
\end{abstract}




\section{Introduction} \label{sec:intro}  
Measured abundance of A(Li) = 3.32 dex in young stars and interstellar medium (ISM) suggest that Li enriched significantly in the Galaxy from primordial value of A(Li) = 2.72~dex, a value predicted from the big bang nucleosynthesis (BBN) models based on the results from WMAP \citep{Cyburt2008}.
One of the known sources for Li enrichment, apart from cosmic ray spallation \citep{HEMitler1972}, an interaction of high energy cosmic ray particles with carbon and oxygen atoms, and novae explosions 
\citep{TajitsuSadakaneNaitoAraiNaito2015, IzzoDellaValleMasonMatteucci2015}, is nucleosynethsis in evolved stars.
Though, stars, in general, are considered as Li sinks, highly evolved stars such as asymptotic giant branch (AGB) stars are known to produce fresh Li \citep{SmithLambert1989} via Cameron-Fowler mechanism \citep{CameronFowler1971} which add to the Li enrichment of the Galaxy through mass loss. It is understood Li production in AGB happens only in relatively massive stars of M $\geq$ 3 M$_{\odot}$ through a process known as hot bottom burning \citep{N.Mowlavi1995, LattanzioFrostCannonWood1996}. However, the discovery of high Li abundance in relatively less evolved low mass red giant branch (RGB) stars was unexpected \citep{WallersteinSneden1982}. Since then, observational studies revealed more Li-rich giants, and systematic surveys suggest Li-rich giants are rare, $<$ 1$\%$  \citep[e.g.,][]{BrownSnedenLambert1989, CharbonnelBalachandran2000, KumarReddy2011}. There are close to 200 Li-rich RGB stars \citep{CaseyRuchtiMasseronRandich2016} known so far with Li $\geq$ 1.5~dex, an upper limit commonly used in the literature and attributed to the standard models of 1st-dredge-up on RGB \citep{Iben1967}. In some cases, Li abundance was found to be more than the star's initial abundance of A(Li) $=$ 3.2~dex, which are known as super Li-rich giants \citep[e.g.,][]{KumarReddy2011, Hong-LiangYanShiZhou2018}. In fact, observations in general, show much less Li than the expected limit \citep{BDFieldsLiProblem2011} and  are attributed to extra-mixing process, post 1st dredge-up, during luminosity bump evolution \citep{DeliyannisCP1990}, which is not part of the standard models. Thus, Li enhancement in RGB giants is a puzzle and it's origin is being debated since a few decades. 
  
There are two main hypotheses proposed for high Li in RGB stars: addition of Li rich material through merger of sub-stellar objects or internal nucleosynthesis and mixing-up of fresh Li with photosphere. There is no consensus on any of these two hypotheses. If the mechanism is in-situ nucleosynthesis as suggested by a few studies \citep{CameronFowler1971, N.Mowlavi1995, LattanzioFrostCannonWood1996}, RGB stars should be counted for Li enrichment in the Galaxy. On the other hand, if the enhancement is due to an external origin, we are simply observing the locked-up Li in sub-stellar objects. Thus, understanding the mechanism of Li enhancement in RGB may not only provide clues to our understanding of RGB nuleosynethesis, but to the larger issue of Li enrichment in the Galaxy.

The clue probably lies in finding the evolutionary phase of Li-rich giants on RGB. Evolutionary phase may be determined either by using star's position in the HR diagram or using asteroseismology \citep{BeddingMosser2011Natur, VrardMosserSamadi2016} or secondary calibrations based on asteroseismology \citep{TingHawkinsRix2018}.
Most of the currently known Li-rich giants have been classified based on their positions in the HR diagram. Due to small separation in values of luminosity and $T_{\rm eff}$ between different phases on RGB, in particular between the red clump and the luminosity bump, there are conflicting reports in the literature about the true evolutionary phase of Li-rich giants  on RGB. With regards to asteroseismic analysis which has been proved to be a gold standard to separate He-core burning giants (red clump) from those of H-core burning giants ascending RGB for the fist time. Asteroseismic data is mostly based on either Kepler or CoRoT missions.  
Unfortunately data is not available for all the stars as these observations are made for specific sky fields. Presently, evolutionary phase is accurately known, based on Kepler asteroseismic data and CoRoT asteroseismic data, only for half a dozen Li-rich stars \citep{SilvaAguirreRuchti2014ApJ, JofreKIC9821622.2015, BharatRaghuReddy2018, RaghuReddyBharat2018, SmiljanicFranciosini2018}. Interestingly, all of them are red clump stars with He-core burning except the star studied by \cite{JofreKIC9821622.2015} which is on RGB with H-core burning phase whose Li-rich giant status is discussed later in the discussion section. None of the previously known Li-rich giants are in the Kepler field. 

This necessitates to explore alternate ways to find the evolutionary phase of Li-rich giants. One of the ways is to undertake large scale survey for Li-rich giants  in globular clusters which have well defined  RGB with visible luminosity bump and red clump with significant number of stars. However, getting a reasonable resolution spectra for a large number of fainter stars in globular clusters even with 10-m class telescopes is difficult and time consuming. Another way is to employ a large data set of giants with measured Li abundance and accurate astrometry.  

Fortunately, while analyzing large data sets from the Gaia and the GALAH spectroscopic surveys to understand elemental abundance patterns in the Galaxy, we noticed a strikingly well populated red clump and RGB in the HR diagram of L-$T_{\rm eff}$ of the GALAH stars. This could, potentially, be used to find clues to Li enhancement origin in RGB stars if we search for Li enhanced giants along the RGB. Combining the Gaia astrometry and the GALAH Li abundances, we provide an evidence that most of the Li-rich stars belong to the red clump  with implications for Li enhancement origin scenarios.

\section{Stellar Sample} \label{sec:Sample}
The Galactic Archeology with HERMES (GALAH) spectrograph is a large scale spectroscopic survey of  spectral resolution of R (=$\Delta \lambda/\lambda$ ) $\approx$ 28,000 with spectrograph attached to 4-m Anglo-Australian Telescope (AAT). 
In the GALAH DR2 (GALAH Data Release 2), the GALAH team has provided quantitatively derived abundances of about 23 elements including the light element Li, which is the focus of this article, for a sample of 342,682 stars \citep[see][]{BuderGalahDR22018}.
The sample is searched for corresponding astrometric and photometric data from the Gaia Data Release 2 (hereafter the Gaia DR2) of 1.7 billion stars.
Source matching is done using the Gaia DR2 source identifier provided in the GALAH DR2 data set \citep[see][]{BuderGalahDR22018, BrownGaiaDr2Summary2018}. This resulted in a sample of 340,299 common stars among the two data sets. Further, we culled out stars for which $\rm \sigma_{Teff_{GALAH}}\geq100\ k$, parallax ($\pi$) is negative and fractional error in parallax, $\rm \sigma_{\pi}$/$\rm \pi$ $\geq$ 0.15 yielding a total sample of 246,390 stars.

The GALAH also provides an index of $\rm Flag_{Cannon}$ against each star which indicates accuracy or level of confidence in derived stellar parameters. $\rm Flag_{Cannon}$ index  range from 0 to 128, with 0 being the most reliable, and other numbers indicating issues such as binary companion, fitting accuracies, S/N ratio, etc. \citep[for more details see][]{BuderGalahDR22018}. We restricted our sample to stars with reliable stellar parameters by culling out all the stars with nonzero value of Flag$_{cannon}$. This yielded a total of 204,370 stars. 

In addition, to avoid stars of higher masses of early AGB and stars evolved from hotter spectral types from main sequence, as some of them are found to be having higher Li abundances due to inefficient mixing, we have restricted our sample to low mass RGB stars of M $\leq$ 2M$_{\odot}$, which are expected to evolve through both luminosity bump and He-flash \citep{CarlbergCunhaSmith2016, CassisiSalarisPietrinferni2016}.
Masses have been estimated using the formula,
{$\rm M/M_\odot=10^{[log(L/L_\odot) +  
log{\it g} - log{\it g_\odot} + 4 \times log(Teff_\odot/Teff)]}$}, 
where we used
log${\it g_\odot}$ = 4.44~dex and Teff$_\odot$ = 5772 K for solar values. Values of effective temperature, Teff$_{\rm GALAH}$, and logarithmic surface gravity, log{\it g}, are adopted from the GALAH catalog \citep{BuderGalahDR22018} and the values of luminosities are based on parallaxes and apparent magnitudes taken from the Gaia catalogue.
We used bolometric correction based on $\rm Teff_{\rm GALAH}$ and relation given in \cite{AndraeFouesneauGaiaDR2StellarParametersFromAPSIS2018}.
Values of luminosities given in the Gaia and those derived in this study using $\rm Teff_{\rm GALAH}$ agrees well with each other. The difference between the two values is quite small with $\sigma\approx$ 0.02~dex.
Uncertainties in the derived luminosity values due to errors in parallax and $\rm Teff_{\rm GALAH}$ are also estimated and are found to be quite small, typically $\sigma_{L/L_\odot}$ $\leq$ 0.05~dex for most of the stars.
With this, we have a final sample of 188,679 low mass stars.

\section {Analysis}

\subsection{Sample stars in L - $T_{\rm eff}$ plane}

\begin{figure}
\includegraphics[width=0.5\textwidth]{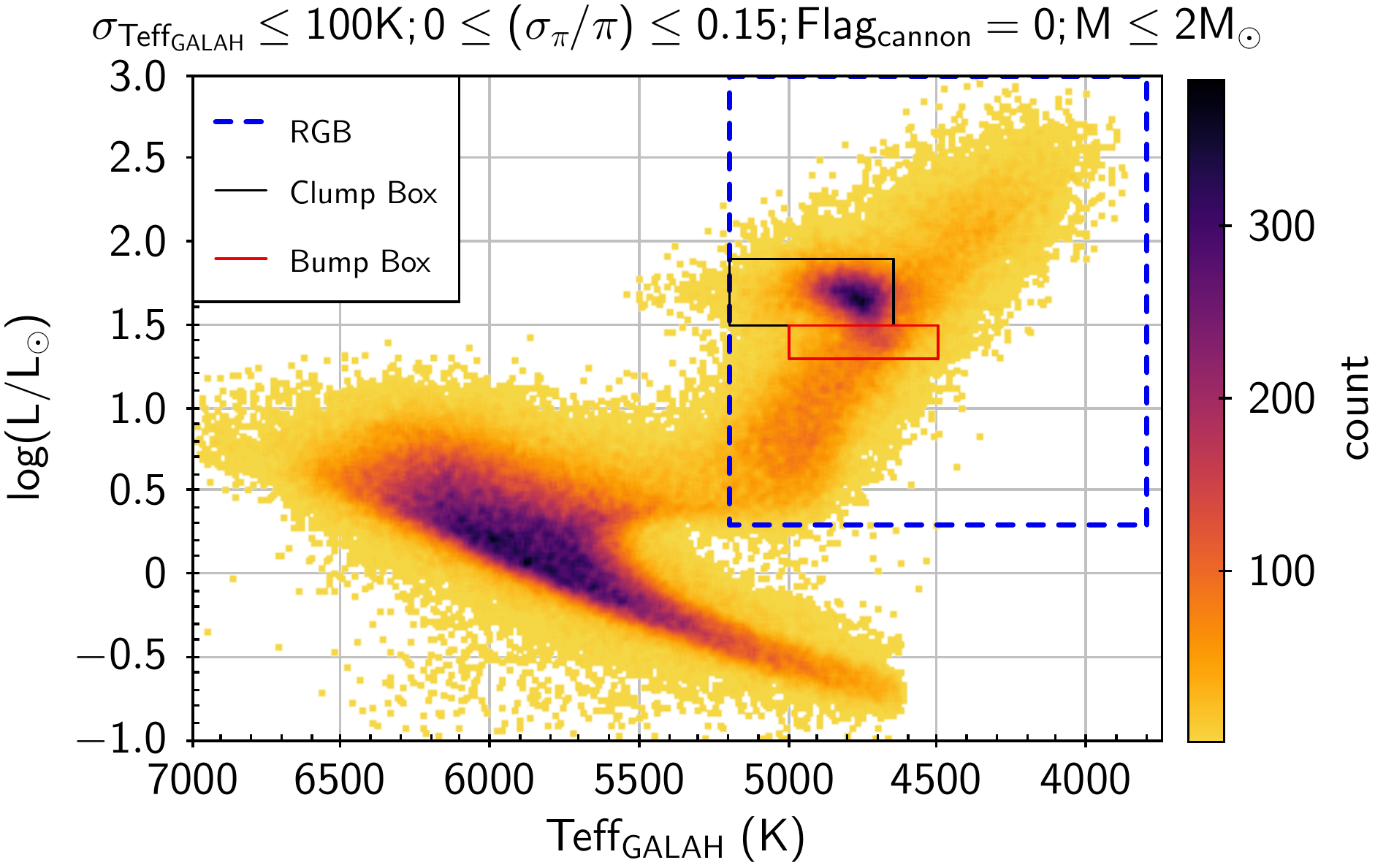}
\caption{ Sample stars in the HR diagram.
\label{fig:LumTeff_Li_All}}
\end{figure}

Sample data plotted in the HR-diagram of Luminosity versus effective temperature ({\color{red} Figure-\ref{fig:LumTeff_Li_All}}) shows stellar evolution with well defined main sequence and red giant branch. Since our interest lies in the Li enhancement during  RGB evolution, RGB has been separated from the data set as defined by a box with dash-line in {\color{red}Figure-\ref{fig:LumTeff_Li_All}}. The box includes giants from the base of RGB through its tip and red clump in the L - $T_{\rm eff}$ plane defined by
 3800K $\leq$ $\rm Teff_{GALAH}$ $\leq$ 5200K and  $0.3 \leq$  $\rm log(L/L_\odot)$ $\leq$ 3.0. This yielded 51,982 red giant branch stars which are the focus of this study. The data sets also show two visible concentrations of stars which are identified as luminosity bump and red clump regions and have been marked by red and black rectangles, respectively. The two clusters are defined in L - $T_{\rm eff}$ plane by (1.5, 1.9; 4650, 5200) for red clump and (1.3, 1.5; 4500, 5000) for the luminosity bump. The concentrations at these two phases are direct consequences of their relatively longer evolutionary timescales compared to any other place on RGB. The well defined stellar concentrations could be used to decipher the origin of Li excess by searching for Li-rich giants among the well defined RGB, luminosity bump and red clump.  

\subsection{Li abundances of sample stars}
The GALAH DR2 provides Li abundances in the form of [Li/Fe] (= [Li/H] - [Fe/H], where [Li/H] = A(Li)- A(Li)$_\odot$) along with associated error and a flag indicating the level of confidence in the measured abundance. The GALAH also provides metallicity, [Fe/H], for all the stars in the catalogue. Commonly, Li abundance is expressed as A(Li) = log (n(Li)/n(H)) + 12. For a better comparison with the literature, [Li/Fe] values of the GALAH are converted into A(Li) using the expression A(Li) = [Li/Fe] + [Fe/H] + A(Li)$_\odot$, where A(Li)$_\odot$ is the solar Li abundance. We adopted A(Li)$_\odot$ = 1.05 $\pm$ 0.10~dex, which is the same one used for the GALAH data \citep[see][]{BuderGalahDR22018}. The corresponding error is $\rm \sigma_{A(Li)}=(\sigma_{[Li/Fe]}^2+\sigma_{[Fe/H]}^2+0.10^2)^{1/2}$. 

Level of confidence in measured lithium abundances of the sample has been encoded in the form of lithium abundance flag, Flag$\rm_{[Li/Fe]}$, against each star.
$\rm Flag_{Abundance}$ index ranges from 0 to 9, with 0 being the most reliable, flag one is raised when line strength is below 2$\sigma$, flag two is raised when cannon fitting started extrapolating, flag three is when both flag one and two are raised, flag four is when $\chi^2$ of best fitting model spectrum is very high or low, flag eight is raised when Flag$\rm_{cannon}$ is not zero, and the remaining flags are combinations of these flags \citep[see][for more details]{BuderGalahDR22018}.
On checking, we found that all the stars with relatively higher value of Li, for example A(Li) $\geq$ 1.6 dex, has Flag$\rm_{[Li/Fe]}$ $\leq$ 3, while for some of the  Li-low stars  Flag$\rm_{[Li/Fe]}$ $\geq$ 3. Higher Flag$\rm_{[Li/Fe]}$ value for Li-low stars is expected as Li absorption line in those stars is weaker. We consider all RGB stars for which Li abundance is measured by the GALAH as this would enable us to comment on the relative fraction of Li-rich giants among RGB stars and evolution of Li itself along the RGB.

\section {Results}

\subsection {Li-rich giants on RGB} \label{sec:LiRichRGB}
Standard models of stellar evolution \citep{Iben1967} predict dilution of Li in RGB stars by a factor of 30 to 60 from its main sequence value. Models, however, do not prescribe the value of initial Li abundance on main sequence which  is a function of age, mass, metallicity, and its pre-main sequence evolution \citep{Lambert&Reddy2004}. If we assume RGB stars evolved off the main sequence with initial value of A(Li) = 3.32~dex, according to the standard models, one would expect A(Li) $\approx$ 1.5 to 1.8~dex for RGB giants in the mass range of 1 to 1.5 M$_{\odot}$, lower the mass, lesser is the dilution factor. However, in the literature, commonly a single upper limit of A(Li) = 1.5~dex is used for Li-normal giants, and giants with A(Li) higher than this are considered as Li-rich.
But for the question ``What is a Li-rich giant?", answer seems to be not a single limit, but probably multiple limits depending on mass and metallicity.
Also, see \cite{SmiljanicFranciosiniRandichMagrini2016} and \cite{AguileraGomezChanamePinsonneaultCarlberg2016} who discussed lower limit for Li-rich giants in connection with two giants (Trumpler 20 MG 340 and 591) with A(Li) $\sim$ 1.6 dex which is at the border line for qualifying them as Li-rich giants.
Here, we use Li trends with  metallicity to define Li abundance limits on RGB. 

\begin{figure}
\includegraphics[width=0.5\textwidth]{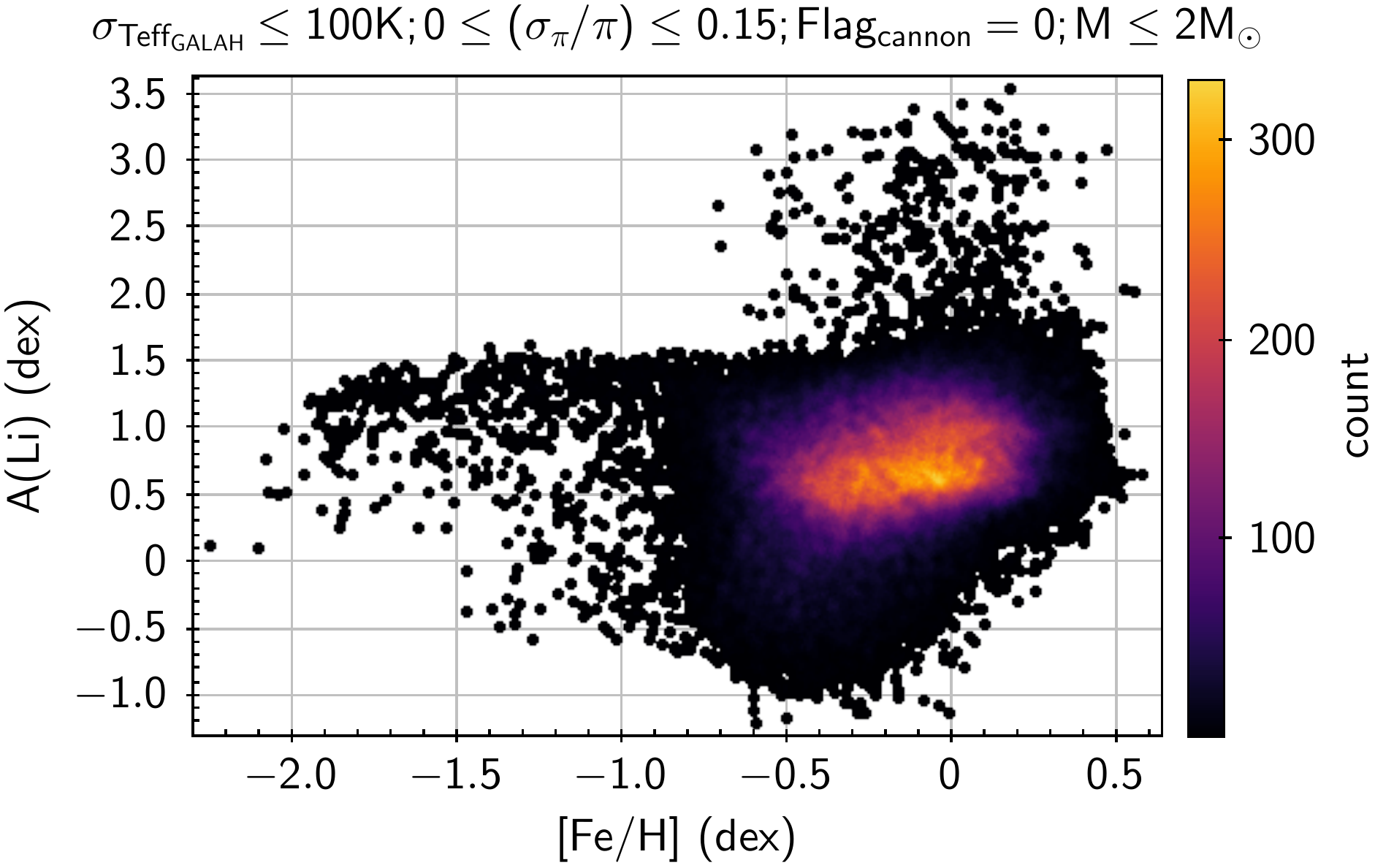}
\caption{Lithium abundance versus metallicity of RGB sample stars.
\label{fig:Li_vs_Fe_FeH_All}}
\end{figure}

\begin{figure*}
\includegraphics[width=1\textwidth]{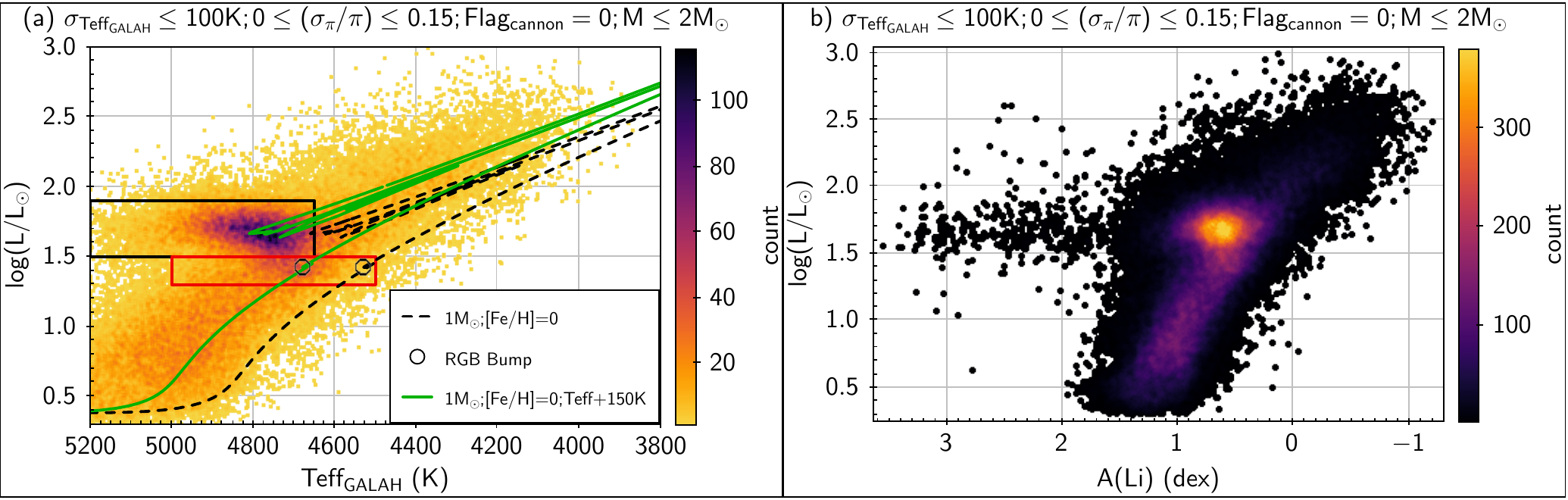}
\caption{The HR diagram and log$\rm (L/L_\odot)$ vs A(Li) distribution for our selected RGB sample stars from the GALAH DR2 and the Gaia DR2 survey with mass M$\leq$2M$_\odot$. In Figure-\ref{fig:LumTeff_Li_RGB_LumTeffandLi}a, we have also shown the approximate positions of RGB bump (red rectangle) and RGB clump (black rectangle).
\label{fig:LumTeff_Li_RGB_LumTeffandLi}}
\end{figure*}

In a plot of A(Li) versus [Fe/H] ({\color{red}Figure-\ref{fig:Li_vs_Fe_FeH_All}}), one can notice varying Li trend with metallicity. The trend can be grouped into three: very metal-poor stars with [Fe/H] $\leq$ $-$1.5 dex, metal poor stars with $-$1.5 $<$ [Fe/H] $\leq$ $-$0.5~dex, and metal-rich stars with [Fe/H] $>$ $-$0.5~dex. The metal-poor giants show a well defined upper limit of A(Li) $\approx$ 1.5 dex for Li-normal giants, which is commonly used in the literature. For very metal poor giants, the nice horizontal trend breaks down at [Fe/H] = $-$1.5~dex and Li abundance starts rapidly decreasing with decreasing [Fe/H].  
Though it is interesting in the context of Li evolution in the early Galaxy, for the present study we concentrate on Li-rich giants among RGB. It is also interesting to note that there are no Li-rich giants among giants with [Fe/H] $\leq$ $-$0.7~dex.
However, on our loosely defined metal-rich side, the Li abundance upper limit for Li-normal stars seems to be slightly higher with a range of A(Li) $\approx$ 1.6 - 1.8~dex than the commonly used upper limit of A(Li) = 1.5~dex for Li-normal giants.   

For the current study, we adopt a single limit of A(Li) = 1.8~dex, instead of using two upper limits, as most of the Li-rich giants in the sample are in the higher metallicity group (i.e [Fe/H] $>$ $-$0.6 dex) as shown in {\color{red}Figure-\ref{fig:Li_vs_Fe_FeH_All}}. This is also the same upper limit set by standard models for low mass (1M$_{\odot}$) Li-normal giants. In addition, given the relatively larger uncertainties ($\approx \pm 0.14$) involved in the derivation of Li abundances from the spectra, adoption of a slightly higher limit to avoid contamination with Li-normal giants is justified. With this definition, we found 335 giants with A(Li) $\geq$ 1.8~dex.
\footnote{Stellar parameters of all the Li-rich giants in our sample will be provided as a supplementary table (online). Currently, data is given at the end of manuscript (see {\color{red} Table-\ref{table:LiRich}}).}
This is about 0.64$\%$ of total RGB giants considered in this study and confirms the previous survey results \citep{BrownSnedenLambert1989, KumarReddy2011, KirbyEvans2012} that Li-rich giants are a rare group.

\subsection{Stellar Evolutionary Phase of Li-rich Giants} 
\label{sec:LiRichRGBEvolutionPhase}
Determination of stellar evolutionary phase of Li-rich giants is the key motivating factor of this study and has been elusive for a long time due to lack of large data set with Li abundances coupled with accurate astrometry. The current larger data set of the GALAH Li abundances and the Gaia astrometry is ideally suited for this purpose. In {\color{red}Figure-\ref{fig:LumTeff_Li_RGB_LumTeffandLi}}, the selected sample of 51,982 giants for which Li abundances have been measured and have accurate $T_{\rm eff}$ and luminosity have been shown. Clump stars are concentrated within the luminosity range of  log(L/L$_{\odot}$) = 1.5 - 1.9 (blue rectabgle box). Bump stars appear to be slightly at a lower luminosity below the much larger group of clump stars. To demonstrate that RGB, bump and clump regions are well identified in the data ({\color{red}Figure-\ref{fig:LumTeff_Li_RGB_LumTeffandLi}}), we superposed a representative theoretical evolutionary track of 1M$_{\odot}$ star with  [Fe/H] = 0.0,
taken from MESA Isochrones \& Stellar Tracks' latest version MISTv1.2 available at \url{http://waps.cfa.harvard.edu/MIST/index.html} and for details see \cite{PaxtonMESACode2011}. The solar like model is used as the mass and metallicity distribution of the current data set peaks at about 1M$_{\odot}$ and [Fe/H] = 0.0 dex. However, the theoretical track is systematically cooler by about 150~K ({\color{red}Figure-\ref{fig:LumTeff_Li_RGB_LumTeffandLi}a}, broken line). If the track is shifted by 150~K towards the hotter side, it coincides well with the observed locations of the bump and clump stars, and runs through the middle of RGB. 
Note, the theoretical models, due to differences in input assumptions, disagree by more than 100 K in $T_{\rm eff}$ among themselves and in some cases with observations \citep{SalarisCassisiWeiss2002}.

To test the occurrence of Li rich giants on RGB, we have shown in {\color{red}Figure-\ref{fig:LumTeff_Li_RGB_LumTeffandLi}b}, a plot of Luminosity versus A(Li). It is known that Li abundance is a function of $T_{\rm eff}$ in which A(Li), in general, decreases with decreasing $T_{\rm eff}$.
As a result, the plot of log(L/L$_{\odot}$) versus A(Li) ({\color{red}Figure-\ref{fig:LumTeff_Li_RGB_LumTeffandLi}b}) mimics the HR diagram of log(L/L$_{\odot}$) verses $\rm Teff_{\rm GALAH}$. Most of the giants follow this general rule and show, on an average, a decreasing Li trend as they ascend RGB towards the tip. Contrary to this general rule, however, a small group of giants show enhanced Li abundance, in particular, at a region which coincides with red clump, post He-flash. There are also Li-rich giants below and above the red clump luminosity range as indicated in {\color{red}Figure-\ref{fig:LumTeff_Li_RGB_LumTeffandLi}a}. Interestingly, Li-rich giants with log(L/L$_{\odot}$) $\leq$1.5 are far less compared to those at the clump. Giants above the clump luminosity are probably early AGB stars about which we comment further in {\color{red} section \ref{sec:Discussion}}.

\begin{figure*}
\includegraphics[width=1\textwidth]{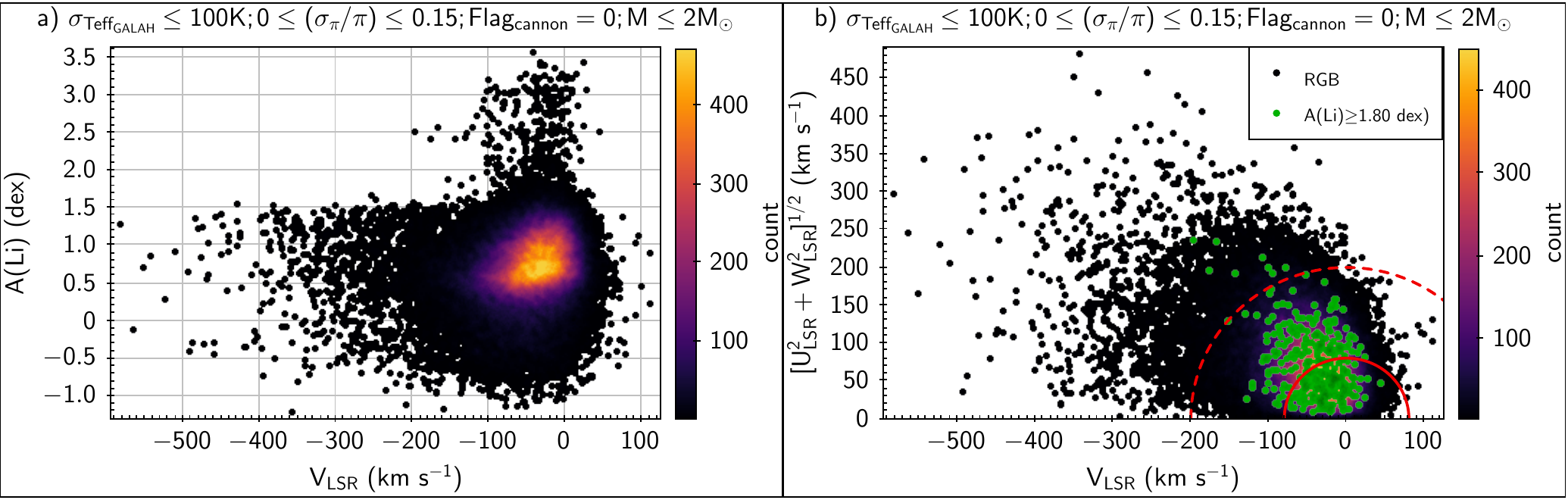}
\caption{Distribution of selected RGB sample in A(Li) vs V$\rm_{LSR}$ plane and in the form of Toomre Diagram. In Figure-\ref{fig:Li_vs_UVW_ToomreDiagram_All}b, we have plotted the approximate boundaries separating thin and think disk (V$\rm_{Total}$=80 km s$^{-1}$, continuous red circle), and thick disk and halo (V$\rm_{Total}$=200 km s$^{-1}$, dotted red circle).
\label{fig:Li_vs_UVW_ToomreDiagram_All}}
\end{figure*}

It appears from {\color{red}Figure-\ref{fig:LumTeff_Li_RGB_LumTeffandLi}} that most of the Li-rich giants are concentrated at the red clump.
Out of the total 335 Li-rich giants, as mentioned above, there are 253 that seem to fall in the red clump luminosity range of L/L$_\odot = 1.5 -1.9$~dex (see {\color{red}Figure-\ref{fig:LumTeff_Li_RGB_LumTeffandLi}a}).
Bump with relatively lesser number of Li-rich giants (24), appears to be just below and rightwards of the clump, falls in the luminosity range of L/L$_\odot\approx 1.3-1.5$~dex.
Given the small difference between the luminosities of red clump and RGB bump, about 0.3 dex (the difference between the mid values), and the typical errors in luminosities, about 0.05 dex (maximum 0.1 dex), we can not rule out the possibility of small overlap of giants among bump and clump.
There are also a few Li-rich giants that are below the bump luminosity about which we comment later.

\subsection{Li-rich Giants in the Galaxy}
To understand the distribution of Li-rich giants among different stellar components of the Galaxy, we have computed membership probability, based on the recipe given in \cite{Reddy2006} and references therein, for each of the stars in the selected RGB sample for belonging to one of the three main components of the Galaxy, namely; thin disk, thick disk and halo.
For this, the heliocentric velocities (U,V,W) for each of the sample giant is calculated using the Gaia astrometry (positions, parallax, and proper motions) and radial velocities (RV) from the GALAH DR2.
The heliocentric velocities are corrected for the solar motion using U$_{o}$=10, V$_{o}$=5.3, W$_{o}$=7.2 (km s$^{-1}$) from \cite{DehnenBinneyLSR1998} to get velocities with respect to the local standard of rest (U$_{\rm LSR}$, V$_{\rm LSR}$, W$_{\rm LSR}$). Entire sample is shown in Toomre Diagram of rotational velocity (V$_{\rm LSR}$) and sum of quadrature of radial and vertical velocities ([$\rm U_{\rm LSR}^2 + W_{\rm LSR}^2]^{1/2}$) \citep{SandageToomreDiagram1987}. Li-rich giants are identified by colored symbols as shown in {\color{red} Figure-\ref{fig:Li_vs_UVW_ToomreDiagram_All}}.
We used kinematic boundaries for thin disk ($\rm |V_{Total}|$ = $\rm [U_{\rm LSR}^2 + V_{\rm LSR}^2 + W_{\rm LSR}^2]^{1/2}$ $\leq$ 80 km s$^{-1}$), thick disk (80 $<|$V$_{\rm Total}| \leq$ 200 km s$^{-1}$) and halo ($|$V$_{\rm Total}| >$ 200 km s$^{-1}$) which are in accordance with the results in \cite{Reddy2006}. Further to quantify the number of giants belonging to different components of the Galaxy,
we used probability of 70$\%$ or more for any star being considered as a member of particular component. We found 223 Li-rich thin disk stars, which is about 0.8$\%$ of total thin disk RGB stars with $P_{\rm thin} \geq 70 \%$ and 69 Li-rich thick disk stars consisting of 0.5$\%$ of total thick disk RGB stars with $P_{thick} \geq 70\%$. We have found just 3  Li-rich giants among 1442 halo giants with $P_{halo} \geq 70\%$. This shows that Li-rich giants are rare ($< 1\%$) across stellar components but are relatively more prevalent among metal-rich thin disk component compared to thick disk and very metal poor halo.

\section{Discussion}\label{sec:Discussion}
Surveys based on large pre-selected low mass RGB stars such as \cite{KumarReddy2011} and \cite{SmiljanicFranciosini2018} using data from Hipparcos and {\it Gaia} catalogue respectively, suggested that Li-rich giants occupy a luminosity range in the HR diagram overlapping the luminosity of RC and RGB bump regions. This led to suggestions of internal nucleosynthesis as both the regions are associated with key phases of stellar evolution: removal of H-profile discontinuity and He-flash, respectively. There are observational reports suggesting existence of  Li-rich stars before and after the luminosity bump on RGB, which indicates that Li-rich giants may occur anywhere on RGB, implying external origin for Li excess such as engulfment of sub-stellar objects. Now the question is whether the Li-rich giants happen anywhere along RGB or at a particular phase on RGB. Answer to this is a key to understand the origin of excess Li in RGB stars.

Thanks to the Kepler and CoRoT missions which provided wealth of time resolved photometric data for stars in certain fields of the sky. Using asteroseismic analysis, one can distinguish RGB giants of H-core burning from those of red clump giants of He-core burning \citep{BeddingMosser2011Natur}. Recently, half a dozen Li-rich giants which are in the Kepler fields haven been reported  by different studies \citep{SilvaAguirreRuchti2014ApJ, BharatRaghuReddy2018, RaghuReddyBharat2018, SmiljanicFranciosini2018}. All of them have been classified as red clump stars with He-core burning based on asteroseismic data analysis with seismic parameters; average period spacing between $g$- and $p$-modes, $\Delta$p $\geq$ 150 and frequency separation, $\Delta \nu$ $\leq$ 5. There is also a lone Li-rich giant (KIC~9821622) which is in Kepler field and has been classified as RGB star with H-core burning \citep{JofreKIC9821622.2015}. Values of luminosity and $T_{\rm eff}$ suggest its location below the luminosity bump on RGB. Though the star is a bona fide RGB, given its reported large difference in derived Li abundance between the stronger resonance line at 6707\AA\ (A(Li)$\rm_{LTE}$ = 1.49 dex; A(Li)$\rm_{NLTE}$ = 1.65 dex) and weaker sub-ordinate line at 6103\AA\ (A(Li)$\rm_{LTE}$ = 1.80 dex; A(Li)$\rm_{NLTE}$ = 1.94 dex), its status being Li-rich giant needs further studies. Also, star's average Li abundance of (A(Li) = 1.80 $\pm$ 0.2~dex for a star with [Fe/H] = $-$0.49 $\pm$ 0.03~dex) is at the border of the limit for Li-normal giants. However, in the absence of Kepler data for most of the known Li-rich stars and lack of systematic surveys for Li-rich giants among asteroseismically known RGB and RC stars, one cannot affirm that Li-rich stars belong only to RC region based on a small number of Li-rich stars randomly discovered. Large data set assembled here is quite suitable to find  likely occurrence of Li-rich giants on different locations on RGB and possible origin mechanisms: external versus in-situ scenarios.

\begin{figure}
\includegraphics[width=0.5\textwidth]{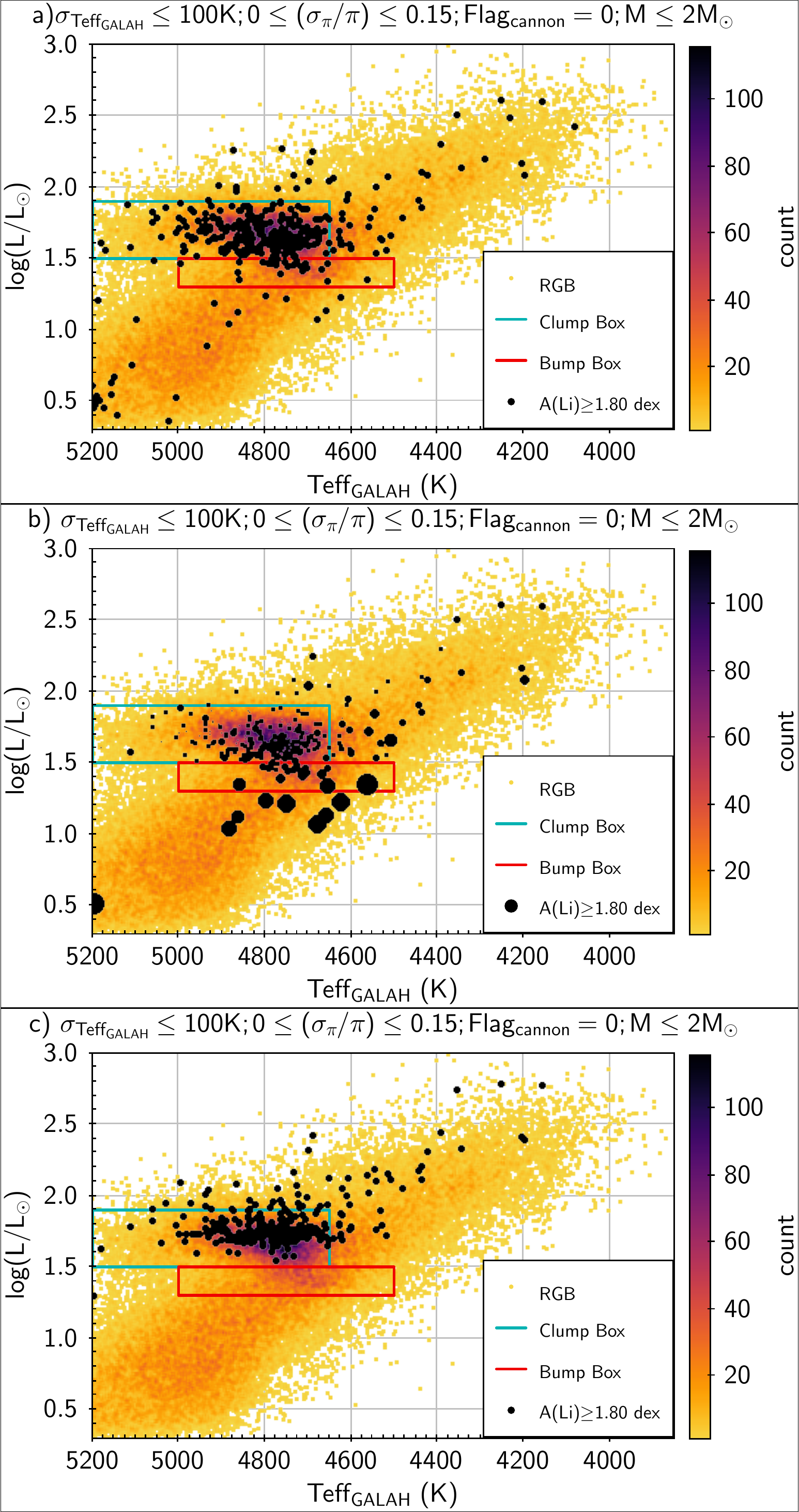}
\caption{The HR diagram showing Li-rich giants' positions a) prior to extinction correction, b) with size of representative point proportional to extinction, and c) post extinction correction.
\label{fig:LiRichFinal}}
\end{figure}

\begin{table}
\caption{Estimated evolutionary timescales for a representative 1M$_{\odot}$ star with metallicities,[Fe/H]=0.0~dex and $-$0.5~dex.
}\label{table:TimeScale}
\begin{tabular}{|c|c|c|c|c|c|c|c|c|c|}
\hline
Evolutionary Phase& For M=1M$_\odot$ \& & For M=1M$_\odot$ \&\\
 & [Fe/H]=0~dex & [Fe/H]=$-$0.5~dex\\
 & t$_e$ (Gyr) & t$_e$ (Gyr)\\
 \hline
MS Turn-off to RGB tip &  1.416 &  0.931 \\
 \hline
RGB base to tip &  0.659  &  0.466 \\
 \hline
RGB base to Bump &  0.555  &  0.402 \\
 \hline
RGB Bump &  0.014  &  0.006 \\
 \hline
RGB Bump to RGB tip &  0.100  &  0.059 \\
\hline
He Core Burning &  0.111  &  0.102 \\
\hline
\end{tabular}
\end{table}

\subsection {External versus in-situ Li enrichment }
We examine external scenario of a planet or sub-stellar material accretion for excess Li in RGB stars. If this scenario has to hold, one would expect Li-rich giants across RGB with relatively higher concentration both at luminosity bump and red clump as their evolutionary timescales are relatively longer. In {\color{red} Table-\ref{table:TimeScale}}, we have given approximate evolutionary periods for different phases of RGB evolution for a representative 1M$_{\odot}$ mass star of metallicity of [Fe/H] = 0.0 dex and $-$0.5 dex.
Timescales are based on theoretical tracks taken from MESA Isochrones \& Stellar Tracks' latest version MISTv1.2 available at \url{http://waps.cfa.harvard.edu/MIST/index.html} and see \cite{PaxtonMESACode2011} for more details.
Evolutionary period for giants between base of RGB and tip of RGB is about 6 times more than that of evolutionary period of giants at the clumps which is about 110 Myrs. This means one should expect a number of Li-rich giants by a similar factor more on RGB in ascending phase including the bump compared to number of Li-rich giants at red clump region. Contrary to this, we found less than one third of the total Li-rich giants (about 106) outside the red clump box (see {\color{red} Figure-\ref{fig:LiRichFinal}a}). Some of these, about 16, are lying just above the red clump box which probably are either evolving off red clump phase towards early AGB or descending from RGB tip.
Quite a few (about 16) appear to be located below L/L$_\odot <$ 0.8 dex,  at the base of the RGB. Probably, these may still be in sub-giant phase and their higher Li abundance may be due to in-sufficient mixing as they might not have experienced fully the 1st dredge-up phase. In this context, we would like to mention the recent discovery of a number of metal-poor sub-giants with Li abundances A(Li) $\geq$ 2.0~dex and in some cases exceeding ISM values \citep{HainingAokiMatsuno2018}. There are also suggestions that Li excess seen in post 1st dredge-up on RGB giants probably originated from sub-giants \citep{KirbyGuhathakurtaZhang2016}, though the origin of high Li in sub giants itself is not known yet.

Though the number of Li-rich stars on RGB are less, it is important to know if they are really on RGB in ascending phase or they are misplaced in the HR diagram due to errors. One of the errors that could be specific to individual giants is extinction, which can affect the stars' brightness and hence its position in the HR diagram. In the present study, we have not taken into account the extinction in deriving luminosities, as the extinction values are not available for majority of the sample stars.  Moreover, the corrections for most of the stars in the sample are small as the stars are nearby ($d <$ 5 kpc) bright giants (with 9 $\leq$ $m_{v}$ $\leq$ 14).  
However, on scrutiny of extinction among Li-rich giants, we found the extinction values only for 269 Li-rich giants out of 335. Interestingly, most of the Li-rich giants that are below the bump have relatively larger extinction values compared to others. 
In {\color{red} Figure-\ref{fig:LiRichFinal}b}, for Li-rich giants with extinction values are shown. Level of extinction is represented by the size of the symbol. If we correct $m_{v}$ values for extinction given in the Gaia DR2, many of the Li-rich giants that are below the bump on RGB move towards brighter clump box ({\color{red} Figure-\ref{fig:LiRichFinal}c}). Some of them also moved out of the clump box towards higher luminosities. Particularly interesting is the fact that there are no Li-rich giants at the bump and also below the bump. Though it is compelling to suggest, based on the current data as shown in {\color{red} Figure-\ref{fig:LiRichFinal}c}, that Li-rich giants belong only to RC, we note
extinction has not been applied to the entire sample as it is not available for majority of the sample. One thing that emerges from {\color{red} Figure-\ref{fig:LiRichFinal}} is that large number of Li-rich giants seems to be at RC even though evolutionary period of RGB is many fold longer than RC giants.

\begin{table*}
\caption{Stellar parameters of Super Li-rich giants with Li abundance of A(Li) $\geq$ 3.2~dex in our sample.}
\label{table:SuperLiRich}
\begin{tabular}{|c|c|c|c|c|c|c|c|c|c|}
\hline
\hline
S.No.&Object ID& Teff & log{\it g} & [Fe/H] & m$_v$ & L/L$_\odot$ & A(Li)\\
 & & (K)& (dex) & (dex)& (mag) & (dex) & (dex)\\
 \hline
1 &Gaia DR2 6423511482552457344 & 4828.68 $\pm$ 58.32 & 2.84 $\pm$ 0.14 & 0.18 $\pm$ 0.05 & 12.15 & 1.56 $\pm$ 0.03 & 3.54 $\pm$ 0.12 \\ 
2 &Gaia DR2 6216747182780840576 & 4773.08 $\pm$ 58.40 & 2.69 $\pm$ 0.15 & 0.12 $\pm$ 0.06 & 12.54 & 1.54 $\pm$ 0.07 & 3.41 $\pm$ 0.12 \\ 
3 &Gaia DR2 3080569351805501824 & 4995.53 $\pm$ 95.89 & 2.60 $\pm$ 0.17 & 0.03 $\pm$ 0.07 & 11.07 & 1.71 $\pm$ 0.04 & 3.41 $\pm$ 0.12 \\ 
4 &Gaia DR2 5920543908525756800 & 4815.52 $\pm$ 76.34 & 2.68 $\pm$ 0.18 & 0.14 $\pm$ 0.07 & 11.71 & 1.54 $\pm$ 0.04 & 3.39 $\pm$ 0.13 \\ 
5 &Gaia DR2 5676420200792553600 & 4854.10 $\pm$ 97.57 & 2.31 $\pm$ 0.16 &-0.11 $\pm$ 0.06 & 12.40 & 1.83 $\pm$ 0.07 & 3.38 $\pm$ 0.12 \\ 
6 &Gaia DR2 6721793108675117440 & 4911.04 $\pm$ 90.19 & 2.45 $\pm$ 0.17 &-0.04 $\pm$ 0.07 & 11.36 & 1.64 $\pm$ 0.04 & 3.33 $\pm$ 0.12 \\ 
7 &Gaia DR2 4488063566731544960 & 4778.94 $\pm$ 86.05 & 2.37 $\pm$ 0.16 &-0.02 $\pm$ 0.06 & 12.34 & 1.52 $\pm$ 0.05 & 3.27 $\pm$ 0.12 \\ 
8 &Gaia DR2 2939800046333110272 & 4985.20 $\pm$ 83.68 & 2.56 $\pm$ 0.18 &-0.13 $\pm$ 0.08 & 12.98 & 1.55 $\pm$ 0.06 & 3.26 $\pm$ 0.13 \\ 
9 &Gaia DR2 4168437628181576192 & 4749.88 $\pm$ 89.58 & 2.71 $\pm$ 0.18 & 0.19 $\pm$ 0.07 & 13.23 & 1.21 $\pm$ 0.06 & 3.26 $\pm$ 0.13 \\ 
10 &Gaia DR2 5229729170925959552 & 5038.38 $\pm$ 70.89 & 2.79 $\pm$ 0.17 &-0.15 $\pm$ 0.07 & 11.13 & 1.65 $\pm$ 0.02 & 3.24 $\pm$ 0.13 \\ 
11 &Gaia DR2 5293680581122445184 & 4832.25 $\pm$ 68.88 & 2.50 $\pm$ 0.15 &-0.01 $\pm$ 0.06 & 12.68 & 1.65 $\pm$ 0.04 & 3.23 $\pm$ 0.12 \\ 
12 &Gaia DR2 5242382659974594688 & 4786.39 $\pm$ 51.84 & 2.80 $\pm$ 0.14 & 0.28 $\pm$ 0.05 & 11.57 & 1.60 $\pm$ 0.03 & 3.23 $\pm$ 0.12 \\ 
13 &Gaia DR2 6721685773156936064 & 4751.60 $\pm$ 58.24 & 2.52 $\pm$ 0.15 & 0.11 $\pm$ 0.06 & 10.87 & 1.68 $\pm$ 0.04 & 3.23 $\pm$ 0.12 \\ 
14 &Gaia DR2 5628302754467688576 & 4868.94 $\pm$ 93.16 & 2.37 $\pm$ 0.17 &-0.20 $\pm$ 0.07 & 13.47 & 1.57 $\pm$ 0.04 & 3.21 $\pm$ 0.13 \\ 
15 &Gaia DR2 3202012502737830784 & 4906.32 $\pm$ 86.59 & 2.42 $\pm$ 0.16 &-0.14 $\pm$ 0.06 & 12.78 & 1.77 $\pm$ 0.11 & 3.21 $\pm$ 0.12 \\ 
16 &Gaia DR2 5460011229840058880 & 4813.14 $\pm$ 50.93 & 2.63 $\pm$ 0.13 & 0.16 $\pm$ 0.05 & 12.17 & 1.70 $\pm$ 0.05 & 3.21 $\pm$ 0.12 \\ 
17 &Gaia DR2 5452473905831060480 & 4711.60 $\pm$ 70.91 & 2.17 $\pm$ 0.15 &-0.30 $\pm$ 0.06 & 12.41 & 1.84 $\pm$ 0.07 & 3.20 $\pm$ 0.12 \\ 
18 &Gaia DR2 6162898261508964992 & 4835.99 $\pm$ 77.44 & 2.58 $\pm$ 0.17 & 0.01 $\pm$ 0.07 & 12.59 & 1.70 $\pm$ 0.12 & 3.20 $\pm$ 0.13 \\ 
19 &Gaia DR2 6779302244026689920 & 4541.30 $\pm$ 69.25 & 2.17 $\pm$ 0.15 &-0.48 $\pm$ 0.06 & 12.02 & 1.99 $\pm$ 0.08 & 3.20 $\pm$ 0.12 \\ 
20 &Gaia DR2 3496188144418768640 & 4776.14 $\pm$ 72.41 & 2.59 $\pm$ 0.16 & 0.04 $\pm$ 0.06 & 12.12 & 1.63 $\pm$ 0.05 & 3.20 $\pm$ 0.12 \\ 
\hline
\end{tabular}
\end{table*}

\begin{figure}
\includegraphics[width=0.5\textwidth]{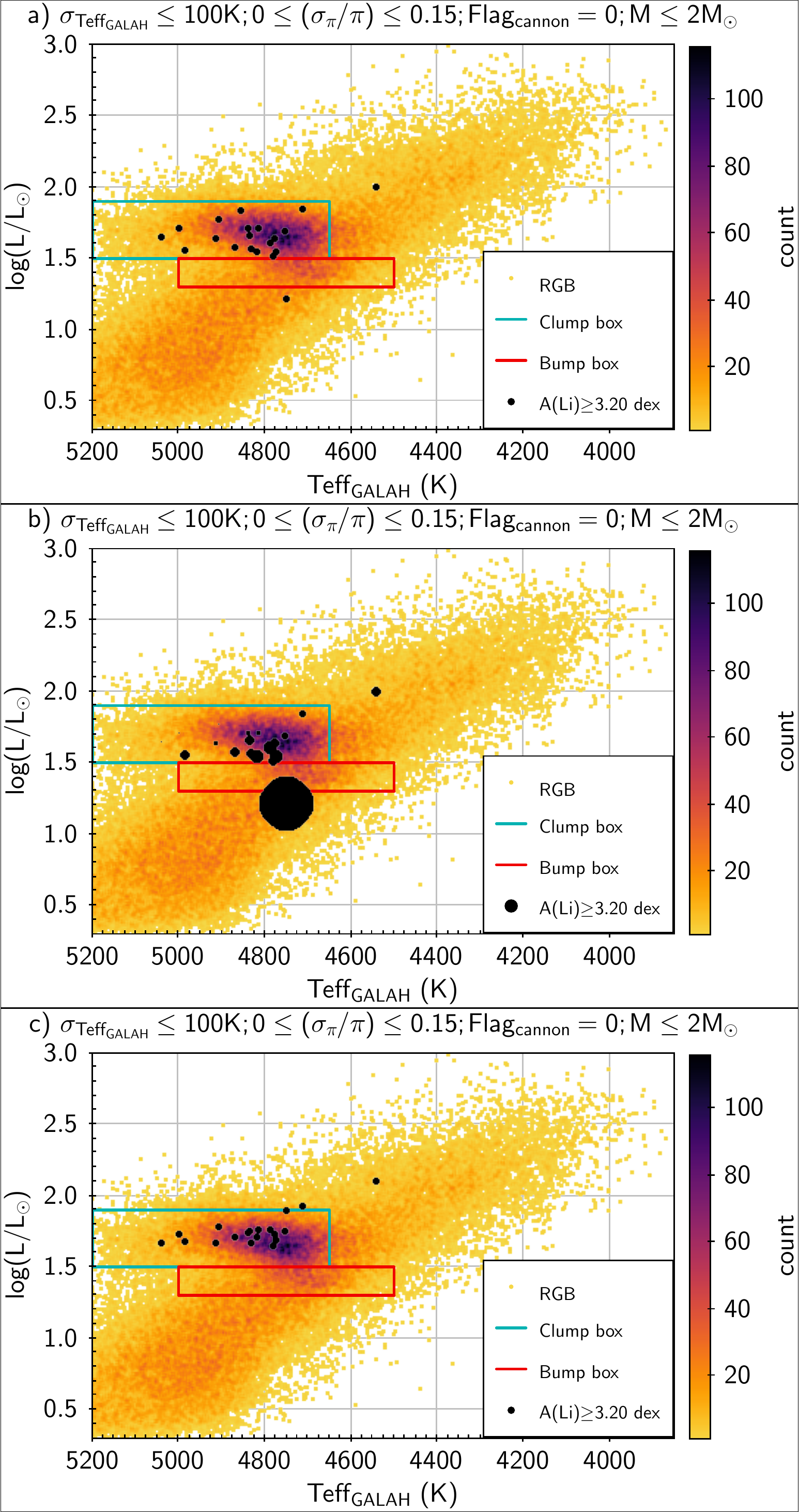}
\caption{The HR diagram showing super Li-rich giants' positions a) prior to extinction correction, b) with size of representative point proportional to extinction, and c) post extinction correction.}
\label{fig:SuperLiRichFinal}
\end{figure}

In {\color{red} Table-\ref{table:SuperLiRich}}, we have separately given super Li-rich giants with A(Li) $\geq$ 3.2~dex. There are 20 of them. Except one super Li-rich giant, all are closer to solar metallicity ([Fe/H] within $\pm 0.30$ dex and belongs to the Galactic thin disk. Super Li-rich giants are plotted in the HR diagram of {\color{red} Figure-\ref{fig:SuperLiRichFinal}a}. With the exemption of two, all falls in RC region defined by turquoise color box. Of two, one is below the bump and the second is above the clump region. Interestingly, the super Li-rich giant that appears below the bump has much larger extinction (larger the symbol higher the extinction) value as shown in {\color{red} Figure-\ref{fig:SuperLiRichFinal}b}. Post extinction correction, the lone super Li-rich giant below the bump moves up into red clump box. After extinction correction, super Li-rich giants seems to fall in the center of  RC region as shown in {\color{red} Figure-\ref{fig:SuperLiRichFinal}c}.
Two of the giants also move slightly upward towards brighter side away from the central region of RC. The giants above the clump are not RGB but are slightly leftwards. Probably, they are either post-clump giants evolving toward early AGB or post He-flash giants.  Most importantly,  
none of the super Li-rich giants are either at  bump or even on RGB. Given the transient nature of Li enrichment and very high Li abundance of super Li-rich giants, the Li enrichment, most probably, would have happened very recently. Clustering of super Li-rich stars only at the clump, as revealed by the data shown in {\color{red} Figure-\ref{fig:SuperLiRichFinal}},  is the most compelling evidence against external scenario for Li-excess origin. There is no reason to believe that the conditions for sub-stellar engulfment at RC are more favorable than at any other RGB phase. This clearly points towards in-situ scenario either at luminosity bump or its immediate preceding phase of He-flash at the RGB tip.

\subsection { In-situ nucleosynthesis at luminosity bump versus He-flash}
 
There are two likely phases at which nucleosynthesis and dredge-up of Li may happen: luminosity bump and  He-flash at the RGB tip. Both the phases are associated with significant internal changes in the stars. It is at the bump at which H-burning shell moves upward and gets encroached by outer convective shell resulting in slowdown of stellar evolution which manifests as luminosity bump \citep{Christensen-Dalsgaard2015}. Also, it is at the bump H-profile discontinuity, a barrier for convective mixing, which gets removed and resumes mixing up of $p-p$- chain products with photosphere. Mixing at the bump, also known as extra-mixing, is attributed for rapid decrease in Li abundance and lowering of carbon isotopic ratios, $^{12}C/^{13}C$ \citep{Gilroy1989, SmiljanicGauderonNorth2009}. On the other hand, He-flash at the tip of RGB terminates RGB phase by removing electron degeneracy in the core \citep{IbenRenzini1984}. Post He-flash, stars settle at red clump quiescently burning He at the core with outer H-burning shell before climbing up towards AGB. 

There are a number of theoretical models dealing with extra mixing at the bump \citep{EggletonDearbornLattanzio2008, CharbonnelLagarde2010, DenissenkovMerryfield2011ApJ...727L...8D, Denissenkov2012ApJ...753L...3D}. The models do suggest synthesis of Li via Cameron-Fowler mechanism in H-shell and somehow dredge-up with photosphere. The models were invoked with the presumption that Li-rich giants occur at the bump as reported by many observational studies \citep{CharbonnelBalachandran2000, KumarReddy2011, Hong-LiangYanShiZhou2018}.  
However, it is not clear how the same mechanism that is responsible for rapid depletion of Li is also responsible for Li enhancement. The lack of clarity on Li origin probably stems from the fact that evolutionary phase of Li-rich giants has not been identified without ambiguity. The reported Li-rich giants fall in a narrow luminosity range which overlaps both bump and clump regions. Difference between clump and bump  values of luminosity and $T_{\rm eff}$  are small, and many a times, they are within the measured uncertainties. This degeneracy even led to the suggestion that Li-rich giants may make zigzag motion and appear as clump or bump due to internal rotation \citep{Denissenkov2012ApJ...753L...3D}. Let us tackle this issue using this large data set and evolutionary period argument, as it requires a large number of stars to define the luminosity bump.

Among 335 Li-rich giants, there are only 106 giants which appear outside the red clump. Only one third of these seem to be either at the bump or above the bump. The large concentration of Li-rich giants at RC suggests, probably, that the origin of Li enhancement lies at or post He-flash rather than at the luminosity bump. If Li enhancement occurred at the bump, the enhanced Li has to be sustained till RC which is quite unlikely given the deep convection post bump evolution. One would also expect significant number of Li-rich giants both at the bump and between bump and the RGB tip, because the evolutionary periods (see {\color{red}Table-\ref{table:TimeScale}}) of giants between the bump to RGB tip (100 Myrs) and the red clump (100 Myrs) are similar. 
This is not the case as shown in {\color{red} Figure-\ref{fig:Li_vs_UVW_ToomreDiagram_All}}, as there is a factor of about four more Li-rich giants at the clump compared to those at the bump or post-bump evolution. One could  make another argument against the bump as the site for Li origin based on very short Li depletion periods compared to evolutionary time of luminosity bump \citep[see][]{PalaciosCharbonnelForestini2001, KumarReddyMuthumariappan2015}. The enhanced Li may deplete within 60$\%$ of the evolutionary period of the bump, i.e Li-rich giants may become Li-normal within a few million years which is much shorter than the evolutionary periods between the bump and RC. 
Based on this  argument, it is reasonable to suggest that all the Li-rich giants are, most likely, in He-core burning phase rather than at multiple  phases on RGB for Li origin. Of course, we need to have further studies to understand a minority of Li-rich giants that are outside of the RC region. 
This conclusion is also strengthened by the fact that all the presently known six Li-rich giants \citep[e.g.,][]{RaghuReddyBharat2018} that are in Kepler field  are red clump giants. Data assembled in {\color{red} Figure-\ref{fig:LiRichFinal}} and {\color{red} Figure-\ref{fig:SuperLiRichFinal}} suggests that most of the Li-rich giants are at red clump which implicitly points towards He-flash, a major stellar event preceding red clump, for Li enrichment. However, we do not have much insight into the nucleosynthesis and dredge-up processes during He-flash, except a couple of studies to understand Li-rich giants at the red clump or the early AGB stars of J-type. For example, the study \cite{MocakMeakinMullerSiess2011} deals with Li production during He-flash, but in case of very metal-poor giants. They show that the requirement of proton injection into the convective shell powered by He burning during He flash may not be met in the case of metal-rich giants as He-ignition occurs relatively farther away from the H-burning shell. For explaining large Li abundances in J-type stars, \cite{ZhangJeffery2013} explored merging up of He white dwarfs (WDs) of different masses with the He-core of RGB giants and showed that they could produce observed values for certain combinations.

\section {Conclusion}

In this paper, we have searched for Li-rich giants among a large data set of red giant branch stars given in the GALAH catalogue  for which the Gaia astrometry is available. We found 335 giants with A(Li) $\geq$ 1.80~dex, an abundance which is more than the maximum Li abundance expected from canonical stellar models for stars in the mass range of 1 to 1.5M$_{\odot}$. Among these, we found 20 super Li-rich giants with A(Li) $\geq$ 3.20~dex. This is the largest number of Li-rich giants ever reported in a single study. Values of Luminosity based on the Gaia parallax and  $T_{\rm eff}$ given in the GALAH suggest that majority of the Li-rich giants are red clump giants as defined in the HR diagram. Only a minority of them are outside it. About a dozen of Li-rich giants are found below the bump which may be sub-giants. The high Li in them may be due to insufficient mixing and dilution from its initial values. There are a couple of Li-rich giants which are above RC and leftwards of RGB, which may either be evolving off the red clump or just on their way to RC after experiencing He-flash at the RGB tip. A few Li-rich giants that seem to fall in the bump region as defined in the HR diagram needs to be studied further.

With the advantage of large data set with well defined evolutionary phases on RGB, we addressed the key issue of origin scenarios: external accretion versus in-situ nucleosynthesis. If the cause for Li excess is due to external accretion, one would expect Li-rich giants across RGB. However, disproportionately large concentration of Li-rich giants at a single evolutionary phase of red clump compared to those on RGB phase, between RGB base and its tip, suggest that external origin for Li enrichment is an unlikely scenario. Though the large data set provides compelling evidence that Li-rich giants may belong to RC with He-core burning phase, we cannot dismiss the possibility of a few Li rich giants on other phases on RGB with different sources of Li enhancement. To firmly establish whether there is single or multiple sites on RGB for Li origin, it is important to conduct further studies such as asteroseismic analysis and detailed abundance studies of all the known Li-rich giants.

\section*{Acknowledgments} \label{sec:acknowledgments}
This work has made use of data from the European Space Agency (ESA) mission {\it Gaia} (\url{https://www.cosmos.esa.int/gaia}), processed by the {\it Gaia} Data Processing and Analysis Consortium (DPAC, \url{https://www.cosmos.esa.int/web/gaia/dpac/consortium}). Funding for the DPAC has been provided by national institutions, in particular the institutions participating in the {\it Gaia} Multilateral Agreement. 
This work has also made use of the GALAH survey which includes data acquired through the Australian Astronomical Observatory. In addition, authors thank Raghubar Singh for his help with stellar evolutionary tracks, and the anonymous reviewers for their constructive comments and suggestions, which helped us to improve the manuscript.

\section*{ORCID iDs}
Deepak: \url{https://orcid.org/0000-0003-2048-9870}\\
B. E. Reddy: \url{https://orcid.org/0000-0001-9246-9743}



\bibliographystyle{mnras}
\bibliography{ref} 







\begin{table*}
\begin{center}
\caption{Stellar parameters of Li-rich giants with Li abundance of A(Li) $\geq$ 1.8~dex in our sample. {\bf (Supplementary table will be available online.)}}\label{table:LiRich}
\end{center}
\begin{tabular}{cccccccccccccc}
\hline
\hline
S.No.&Object ID& Teff & log{\it g} & [Fe/H] & m$_v$ & L/L$_\odot$ & A(Li)\\
 & & (K)& (dex) & (dex)& (mag) & (dex) & (dex)\\
\hline
\hline
1 & Gaia DR2 6423511482552457344 & 4828.68 $\pm$ 58.32 & 2.84 $\pm$ 0.14 & 0.18 $\pm$ 0.05 & 12.15 & 1.56 $\pm$ 0.03 & 3.54 $\pm$ 0.12 \\ 
2 & Gaia DR2 6216747182780840576 & 4773.08 $\pm$ 58.40 & 2.69 $\pm$ 0.15 & 0.12 $\pm$ 0.06 & 12.54 & 1.54 $\pm$ 0.07 & 3.41 $\pm$ 0.12 \\ 
3 & Gaia DR2 3080569351805501824 & 4995.53 $\pm$ 95.89 & 2.60 $\pm$ 0.17 & 0.03 $\pm$ 0.07 & 11.07 & 1.71 $\pm$ 0.04 & 3.41 $\pm$ 0.12 \\ 
4 & Gaia DR2 5920543908525756800 & 4815.52 $\pm$ 76.34 & 2.68 $\pm$ 0.18 & 0.14 $\pm$ 0.07 & 11.71 & 1.54 $\pm$ 0.04 & 3.39 $\pm$ 0.13 \\ 
5 & Gaia DR2 5676420200792553600 & 4854.10 $\pm$ 97.57 & 2.31 $\pm$ 0.16 &-0.11 $\pm$ 0.06 & 12.40 & 1.83 $\pm$ 0.07 & 3.38 $\pm$ 0.12 \\ 
6 & Gaia DR2 6721793108675117440 & 4911.04 $\pm$ 90.19 & 2.45 $\pm$ 0.17 &-0.04 $\pm$ 0.07 & 11.36 & 1.64 $\pm$ 0.04 & 3.33 $\pm$ 0.12 \\ 
7 & Gaia DR2 4488063566731544960 & 4778.94 $\pm$ 86.05 & 2.37 $\pm$ 0.16 &-0.02 $\pm$ 0.06 & 12.34 & 1.52 $\pm$ 0.05 & 3.27 $\pm$ 0.12 \\ 
8 & Gaia DR2 2939800046333110272 & 4985.20 $\pm$ 83.68 & 2.56 $\pm$ 0.18 &-0.13 $\pm$ 0.08 & 12.98 & 1.55 $\pm$ 0.06 & 3.26 $\pm$ 0.13 \\ 
9 & Gaia DR2 4168437628181576192 & 4749.88 $\pm$ 89.58 & 2.71 $\pm$ 0.18 & 0.19 $\pm$ 0.07 & 13.23 & 1.21 $\pm$ 0.06 & 3.26 $\pm$ 0.13 \\ 
10 & Gaia DR2 5229729170925959552 & 5038.38 $\pm$ 70.89 & 2.79 $\pm$ 0.17 &-0.15 $\pm$ 0.07 & 11.13 & 1.65 $\pm$ 0.02 & 3.24 $\pm$ 0.13 \\ 
11 & Gaia DR2 5293680581122445184 & 4832.25 $\pm$ 68.88 & 2.50 $\pm$ 0.15 &-0.01 $\pm$ 0.06 & 12.68 & 1.65 $\pm$ 0.04 & 3.23 $\pm$ 0.12 \\ 
12 & Gaia DR2 5242382659974594688 & 4786.39 $\pm$ 51.84 & 2.80 $\pm$ 0.14 & 0.28 $\pm$ 0.05 & 11.57 & 1.60 $\pm$ 0.03 & 3.23 $\pm$ 0.12 \\ 
13 & Gaia DR2 6721685773156936064 & 4751.60 $\pm$ 58.24 & 2.52 $\pm$ 0.15 & 0.11 $\pm$ 0.06 & 10.87 & 1.68 $\pm$ 0.04 & 3.23 $\pm$ 0.12 \\ 
14 & Gaia DR2 5628302754467688576 & 4868.94 $\pm$ 93.16 & 2.37 $\pm$ 0.17 &-0.20 $\pm$ 0.07 & 13.47 & 1.57 $\pm$ 0.04 & 3.21 $\pm$ 0.13 \\ 
15 & Gaia DR2 3202012502737830784 & 4906.32 $\pm$ 86.59 & 2.42 $\pm$ 0.16 &-0.14 $\pm$ 0.06 & 12.78 & 1.77 $\pm$ 0.11 & 3.21 $\pm$ 0.12 \\ 
16 & Gaia DR2 5460011229840058880 & 4813.14 $\pm$ 50.93 & 2.63 $\pm$ 0.13 & 0.16 $\pm$ 0.05 & 12.17 & 1.70 $\pm$ 0.05 & 3.21 $\pm$ 0.12 \\ 
17 & Gaia DR2 5452473905831060480 & 4711.60 $\pm$ 70.91 & 2.17 $\pm$ 0.15 &-0.30 $\pm$ 0.06 & 12.41 & 1.84 $\pm$ 0.07 & 3.20 $\pm$ 0.12 \\ 
18 & Gaia DR2 6162898261508964992 & 4835.99 $\pm$ 77.44 & 2.58 $\pm$ 0.17 & 0.01 $\pm$ 0.07 & 12.59 & 1.70 $\pm$ 0.12 & 3.20 $\pm$ 0.13 \\ 
19 & Gaia DR2 6779302244026689920 & 4541.30 $\pm$ 69.25 & 2.17 $\pm$ 0.15 &-0.48 $\pm$ 0.06 & 12.02 & 1.99 $\pm$ 0.08 & 3.20 $\pm$ 0.12 \\ 
20 & Gaia DR2 3496188144418768640 & 4776.14 $\pm$ 72.41 & 2.59 $\pm$ 0.16 & 0.04 $\pm$ 0.06 & 12.12 & 1.63 $\pm$ 0.05 & 3.20 $\pm$ 0.12 \\ 
21 & Gaia DR2 5446401096954011264 & 4834.50 $\pm$ 77.10 & 2.27 $\pm$ 0.16 &-0.24 $\pm$ 0.06 & 13.33 & 1.86 $\pm$ 0.08 & 3.18 $\pm$ 0.12 \\ 
22 & Gaia DR2 5380617896081668608 & 4936.61 $\pm$ 82.12 & 2.32 $\pm$ 0.17 &-0.26 $\pm$ 0.07 & 12.95 & 1.77 $\pm$ 0.09 & 3.18 $\pm$ 0.12 \\ 
23 & Gaia DR2 5467139329361254144 & 4683.37 $\pm$ 76.26 & 2.54 $\pm$ 0.17 & 0.20 $\pm$ 0.07 & 12.96 & 1.66 $\pm$ 0.07 & 3.15 $\pm$ 0.13 \\ 
24 & Gaia DR2 5819295071037261824 & 4837.62 $\pm$ 95.07 & 2.60 $\pm$ 0.18 & 0.06 $\pm$ 0.08 & 11.96 & 1.60 $\pm$ 0.03 & 3.10 $\pm$ 0.13 \\ 
25 & Gaia DR2 6385336541913936512 & 4738.25 $\pm$ 82.88 & 1.89 $\pm$ 0.16 &-0.59 $\pm$ 0.06 & 12.28 & 1.84 $\pm$ 0.05 & 3.09 $\pm$ 0.12 \\ 
26 & Gaia DR2 6204948465240692736 & 4791.82 $\pm$ 63.13 & 2.66 $\pm$ 0.17 & 0.15 $\pm$ 0.07 & 12.62 & 1.62 $\pm$ 0.07 & 3.09 $\pm$ 0.12 \\ 
27 & Gaia DR2 5779908846543414016 & 4806.52 $\pm$ 66.85 & 2.56 $\pm$ 0.17 & 0.02 $\pm$ 0.07 & 12.42 & 1.54 $\pm$ 0.04 & 3.09 $\pm$ 0.12 \\ 
28 & Gaia DR2 4140059031209507456 & 4677.78 $\pm$ 66.24 & 3.07 $\pm$ 0.17 & 0.47 $\pm$ 0.07 & 12.97 & 1.07 $\pm$ 0.04 & 3.08 $\pm$ 0.13 \\ 
29 & Gaia DR2 6093873292346168064 & 4802.64 $\pm$ 79.08 & 2.45 $\pm$ 0.19 & 0.05 $\pm$ 0.08 & 11.33 & 1.47 $\pm$ 0.06 & 3.08 $\pm$ 0.13 \\ 
30 & Gaia DR2 6845853685693093504 & 4904.20 $\pm$ 95.05 & 2.16 $\pm$ 0.18 &-0.38 $\pm$ 0.07 & 12.23 & 2.01 $\pm$ 0.09 & 3.07 $\pm$ 0.13 \\ 
31 & Gaia DR2 5623835163845080576 & 4821.88 $\pm$ 65.10 & 2.62 $\pm$ 0.17 & 0.15 $\pm$ 0.07 & 11.63 & 1.56 $\pm$ 0.03 & 3.07 $\pm$ 0.13 \\ 
32 & Gaia DR2 5907710473221745152 & 4797.55 $\pm$ 82.93 & 2.50 $\pm$ 0.19 & 0.10 $\pm$ 0.08 & 12.24 & 1.74 $\pm$ 0.06 & 3.06 $\pm$ 0.13 \\ 
33 & Gaia DR2 4248247505054982144 & 4897.40 $\pm$ 93.05 & 2.82 $\pm$ 0.21 & 0.21 $\pm$ 0.09 & 11.31 & 1.56 $\pm$ 0.05 & 3.06 $\pm$ 0.14 \\ 
34 & Gaia DR2 5776389997014959232 & 4545.21 $\pm$ 94.95 & 2.13 $\pm$ 0.19 &-0.42 $\pm$ 0.08 & 13.82 & 1.84 $\pm$ 0.06 & 3.05 $\pm$ 0.13 \\ 
35 & Gaia DR2 5421808423456535936 & 4838.15 $\pm$ 96.19 & 2.38 $\pm$ 0.18 &-0.13 $\pm$ 0.07 & 12.55 & 1.86 $\pm$ 0.06 & 3.05 $\pm$ 0.13 \\ 
36 & Gaia DR2 5433425549777634304 & 4888.15 $\pm$ 92.06 & 2.53 $\pm$ 0.18 &-0.12 $\pm$ 0.08 & 12.60 & 1.69 $\pm$ 0.05 & 3.05 $\pm$ 0.13 \\ 
37 & Gaia DR2 6093817968865392768 & 5177.02 $\pm$ 83.79 & 2.82 $\pm$ 0.19 &-0.22 $\pm$ 0.08 & 12.46 & 1.60 $\pm$ 0.09 & 3.05 $\pm$ 0.13 \\ 
38 & Gaia DR2 6188759423532409472 & 4674.71 $\pm$ 64.71 & 2.60 $\pm$ 0.16 & 0.16 $\pm$ 0.06 & 12.45 & 1.62 $\pm$ 0.05 & 3.05 $\pm$ 0.12 \\ 
39 & Gaia DR2 5908117846580723328 & 4602.58 $\pm$ 87.89 & 2.11 $\pm$ 0.18 &-0.13 $\pm$ 0.07 & 11.18 & 1.73 $\pm$ 0.03 & 3.05 $\pm$ 0.13 \\ 
40 & Gaia DR2 6703523455968541568 & 4658.06 $\pm$ 86.01 & 2.63 $\pm$ 0.19 & 0.19 $\pm$ 0.08 & 13.57 & 1.69 $\pm$ 0.08 & 3.04 $\pm$ 0.13 \\ 
41 & Gaia DR2 6892346397434608384 & 4698.31 $\pm$ 80.33 & 2.49 $\pm$ 0.18 & 0.04 $\pm$ 0.07 & 12.68 & 1.67 $\pm$ 0.07 & 3.04 $\pm$ 0.13 \\ 
42 & Gaia DR2 6715209443503275776 & 4679.61 $\pm$ 84.15 & 2.31 $\pm$ 0.16 &-0.14 $\pm$ 0.06 & 11.20 & 1.77 $\pm$ 0.06 & 3.04 $\pm$ 0.12 \\ 
43 & Gaia DR2 6014097882591550976 & 4761.92 $\pm$ 88.93 & 2.31 $\pm$ 0.17 &-0.25 $\pm$ 0.07 & 13.57 & 1.39 $\pm$ 0.06 & 3.03 $\pm$ 0.13 \\ 
44 & Gaia DR2 5819154642796257664 & 4702.46 $\pm$ 77.61 & 2.84 $\pm$ 0.20 & 0.32 $\pm$ 0.08 & 11.52 & 1.46 $\pm$ 0.03 & 3.03 $\pm$ 0.14 \\ 
45 & Gaia DR2 5821626207499355392 & 4772.56 $\pm$ 76.32 & 2.63 $\pm$ 0.19 & 0.12 $\pm$ 0.08 & 11.43 & 1.45 $\pm$ 0.03 & 3.03 $\pm$ 0.13 \\ 
46 & Gaia DR2 5634364362072360064 & 4716.95 $\pm$ 87.81 & 2.61 $\pm$ 0.19 & 0.08 $\pm$ 0.08 & 13.63 & 1.57 $\pm$ 0.05 & 3.02 $\pm$ 0.13 \\ 
47 & Gaia DR2 6087493341765160576 & 5183.81 $\pm$ 70.54 & 3.32 $\pm$ 0.19 & 0.23 $\pm$ 0.08 & 13.94 & 1.20 $\pm$ 0.06 & 3.02 $\pm$ 0.13 \\ 
48 & Gaia DR2 6566057220857377664 & 4701.72 $\pm$ 85.32 & 2.10 $\pm$ 0.19 &-0.47 $\pm$ 0.08 & 12.21 & 1.71 $\pm$ 0.09 & 3.02 $\pm$ 0.13 \\ 
49 & Gaia DR2 6130099825359747584 & 4783.58 $\pm$ 55.53 & 2.77 $\pm$ 0.15 & 0.25 $\pm$ 0.06 & 12.14 & 1.62 $\pm$ 0.05 & 3.02 $\pm$ 0.12 \\ 
50 & Gaia DR2 5380837493466374272 & 4781.18 $\pm$ 56.09 & 2.89 $\pm$ 0.16 & 0.39 $\pm$ 0.06 & 12.46 & 1.49 $\pm$ 0.04 & 3.02 $\pm$ 0.12 \\
51 & Gaia DR2 6705449899116253312 & 4883.28 $\pm$ 76.61 & 2.38 $\pm$ 0.19 &-0.11 $\pm$ 0.08 & 11.96 & 1.58 $\pm$ 0.06 & 3.02 $\pm$ 0.13 \\ 
52 & Gaia DR2 5229986692861011200 & 4830.59 $\pm$ 76.71 & 2.62 $\pm$ 0.18 &-0.03 $\pm$ 0.07 & 13.31 & 1.67 $\pm$ 0.04 & 3.01 $\pm$ 0.13 \\ 
53 & Gaia DR2 5703002509304050944 & 4738.37 $\pm$ 79.54 & 2.41 $\pm$ 0.16 &-0.15 $\pm$ 0.07 & 12.98 & 1.77 $\pm$ 0.08 & 3.01 $\pm$ 0.12 \\ 
54 & Gaia DR2 5414135481563472640 & 4744.30 $\pm$ 55.03 & 2.72 $\pm$ 0.16 & 0.25 $\pm$ 0.06 & 11.62 & 1.65 $\pm$ 0.04 & 3.01 $\pm$ 0.12 \\ 
55 & Gaia DR2 2898121301434595584 & 4815.25 $\pm$ 67.73 & 2.39 $\pm$ 0.15 &-0.14 $\pm$ 0.06 & 12.17 & 1.71 $\pm$ 0.04 & 3.00 $\pm$ 0.12 \\
56 & Gaia DR2 5294650522177685120 & 4845.01 $\pm$ 65.21 & 2.55 $\pm$ 0.15 &-0.08 $\pm$ 0.06 & 12.82 & 1.52 $\pm$ 0.03 & 2.98 $\pm$ 0.12 \\ 
57 & Gaia DR2 5823358488074533120 & 4965.30 $\pm$ 96.45 & 2.51 $\pm$ 0.20 &-0.01 $\pm$ 0.08 & 12.14 & 1.64 $\pm$ 0.04 & 2.97 $\pm$ 0.14 \\ 
58 & Gaia DR2 5947597289051638784 & 4763.65 $\pm$ 80.92 & 2.35 $\pm$ 0.18 &-0.07 $\pm$ 0.07 & 12.12 & 1.65 $\pm$ 0.06 & 2.97 $\pm$ 0.13 \\ 
59 & Gaia DR2 6088362363965385216 & 4862.10 $\pm$ 55.61 & 2.42 $\pm$ 0.13 &-0.06 $\pm$ 0.05 & 12.48 & 1.65 $\pm$ 0.07 & 2.96 $\pm$ 0.12 \\ 
60 & Gaia DR2 5572621698926229120 & 4802.91 $\pm$ 85.70 & 2.36 $\pm$ 0.17 &-0.14 $\pm$ 0.07 & 13.21 & 1.66 $\pm$ 0.04 & 2.95 $\pm$ 0.13 \\ 
\hline
\end{tabular}
\end{table*}

\begin{table*}
\begin{center}
\contcaption{} \end{center}
\begin{tabular}{cccccccccccccc}
\hline
\hline
S.No.&Object ID& Teff & log{\it g} & [Fe/H] & m$_v$ & L/L$_\odot$ & A(Li)\\
 & & (K)& (dex) & (dex)& (mag) & (dex) & (dex)\\
\hline
\hline
61 & Gaia DR2 6661697548308365312 & 4660.75 $\pm$ 60.36 & 2.46 $\pm$ 0.17 & 0.09 $\pm$ 0.07 & 12.43 & 1.68 $\pm$ 0.09 & 2.95 $\pm$ 0.12 \\ 
62 & Gaia DR2 4242037738259542784 & 4864.57 $\pm$ 79.32 & 2.35 $\pm$ 0.18 &-0.06 $\pm$ 0.07 & 11.94 & 1.63 $\pm$ 0.05 & 2.95 $\pm$ 0.13 \\ 
63 & Gaia DR2 5204530013679857280 & 4892.15 $\pm$ 88.58 & 2.40 $\pm$ 0.18 &-0.16 $\pm$ 0.07 & 13.24 & 1.54 $\pm$ 0.04 & 2.93 $\pm$ 0.13 \\ 
64 & Gaia DR2 5921483333826980224 & 4809.35 $\pm$ 55.37 & 2.40 $\pm$ 0.12 &-0.09 $\pm$ 0.05 & 12.12 & 1.58 $\pm$ 0.06 & 2.93 $\pm$ 0.11 \\ 
65 & Gaia DR2 5468677168171016320 & 4718.90 $\pm$ 54.08 & 2.48 $\pm$ 0.14 &-0.11 $\pm$ 0.05 & 12.26 & 1.75 $\pm$ 0.06 & 2.92 $\pm$ 0.12 \\ 
66 & Gaia DR2 5814033255067021312 & 4342.28 $\pm$ 91.66 & 1.56 $\pm$ 0.19 &-0.50 $\pm$ 0.08 & 12.64 & 2.13 $\pm$ 0.13 & 2.91 $\pm$ 0.13 \\ 
67 & Gaia DR2 5702810159189857408 & 4730.05 $\pm$ 72.93 & 2.32 $\pm$ 0.16 &-0.02 $\pm$ 0.06 & 13.05 & 1.51 $\pm$ 0.07 & 2.91 $\pm$ 0.12 \\ 
68 & Gaia DR2 5793668547375510016 & 4759.90 $\pm$ 66.51 & 2.13 $\pm$ 0.17 & 0.00 $\pm$ 0.07 & 11.13 & 2.26 $\pm$ 0.04 & 2.91 $\pm$ 0.13 \\ 
69 & Gaia DR2 5246928075402201216 & 4598.41 $\pm$ 60.79 & 2.73 $\pm$ 0.17 & 0.26 $\pm$ 0.07 & 11.39 & 1.56 $\pm$ 0.02 & 2.91 $\pm$ 0.12 \\ 
70 & Gaia DR2 3135783149258877824 & 4917.72 $\pm$ 63.69 & 2.27 $\pm$ 0.16 &-0.55 $\pm$ 0.06 & 12.61 & 1.78 $\pm$ 0.08 & 2.89 $\pm$ 0.12 \\ 
71 & Gaia DR2 6008102726729861632 & 4880.12 $\pm$ 92.64 & 2.87 $\pm$ 0.19 & 0.13 $\pm$ 0.08 & 13.57 & 1.04 $\pm$ 0.35 & 2.89 $\pm$ 0.13 \\ 
72 & Gaia DR2 2679552786563683328 & 4668.91 $\pm$ 71.58 & 2.31 $\pm$ 0.16 &-0.13 $\pm$ 0.06 & 12.91 & 1.77 $\pm$ 0.10 & 2.88 $\pm$ 0.12 \\ 
73 & Gaia DR2 6736616969290744576 & 4789.81 $\pm$ 68.49 & 2.56 $\pm$ 0.17 & 0.13 $\pm$ 0.07 & 10.91 & 1.60 $\pm$ 0.05 & 2.88 $\pm$ 0.12 \\ 
74 & Gaia DR2 6688605041679237248 & 4887.88 $\pm$ 69.33 & 2.36 $\pm$ 0.16 &-0.32 $\pm$ 0.06 & 12.59 & 1.68 $\pm$ 0.07 & 2.86 $\pm$ 0.12 \\ 
75 & Gaia DR2 4486423365960888192 & 4731.80 $\pm$ 45.68 & 2.28 $\pm$ 0.11 &-0.15 $\pm$ 0.04 & 11.92 & 1.60 $\pm$ 0.04 & 2.86 $\pm$ 0.11 \\ 
76 & Gaia DR2 5371457559771370368 & 4809.21 $\pm$ 60.44 & 2.65 $\pm$ 0.17 & 0.16 $\pm$ 0.07 & 11.51 & 1.66 $\pm$ 0.06 & 2.86 $\pm$ 0.12 \\ 
77 & Gaia DR2 6735047343387113088 & 4736.17 $\pm$ 87.03 & 2.78 $\pm$ 0.19 & 0.20 $\pm$ 0.08 & 13.41 & 1.58 $\pm$ 0.06 & 2.85 $\pm$ 0.13 \\ 
78 & Gaia DR2 6094984761160506240 & 5057.13 $\pm$ 55.82 & 2.75 $\pm$ 0.15 &-0.12 $\pm$ 0.06 & 10.73 & 1.75 $\pm$ 0.05 & 2.85 $\pm$ 0.12 \\ 
79 & Gaia DR2 6082878897620910336 & 4832.60 $\pm$ 85.91 & 2.73 $\pm$ 0.19 & 0.22 $\pm$ 0.08 & 10.66 & 1.66 $\pm$ 0.07 & 2.85 $\pm$ 0.13 \\ 
80 & Gaia DR2 6654103668171738880 & 4867.99 $\pm$ 68.80 & 2.53 $\pm$ 0.16 &-0.07 $\pm$ 0.07 & 12.67 & 1.73 $\pm$ 0.07 & 2.84 $\pm$ 0.12 \\ 
81 & Gaia DR2 6134488083643500928 & 4842.17 $\pm$ 58.42 & 2.47 $\pm$ 0.15 &-0.06 $\pm$ 0.06 & 12.28 & 1.67 $\pm$ 0.07 & 2.84 $\pm$ 0.12 \\ 
82 & Gaia DR2 5661751169489845760 & 4759.45 $\pm$ 76.23 & 2.11 $\pm$ 0.17 &-0.34 $\pm$ 0.07 & 13.19 & 1.71 $\pm$ 0.13 & 2.82 $\pm$ 0.13 \\ 
83 & Gaia DR2 5790803112691487232 & 4712.54 $\pm$ 68.05 & 2.64 $\pm$ 0.17 & 0.28 $\pm$ 0.07 & 13.81 & 1.44 $\pm$ 0.04 & 2.82 $\pm$ 0.13 \\ 
84 & Gaia DR2 5367726126545882240 & 4436.05 $\pm$ 59.71 & 2.18 $\pm$ 0.16 &-0.16 $\pm$ 0.06 & 11.36 & 2.10 $\pm$ 0.04 & 2.82 $\pm$ 0.12 \\ 
85 & Gaia DR2 4359923561049343744 & 4652.36 $\pm$ 77.52 & 2.96 $\pm$ 0.21 & 0.39 $\pm$ 0.09 & 13.33 & 1.33 $\pm$ 0.07 & 2.82 $\pm$ 0.14 \\ 
86 & Gaia DR2 4904707520793305216 & 4730.07 $\pm$ 90.88 & 2.31 $\pm$ 0.19 &-0.19 $\pm$ 0.08 & 12.96 & 1.71 $\pm$ 0.06 & 2.80 $\pm$ 0.13 \\ 
87 & Gaia DR2 5470340896766060800 & 4752.70 $\pm$ 73.82 & 2.62 $\pm$ 0.19 & 0.20 $\pm$ 0.08 & 12.87 & 1.63 $\pm$ 0.08 & 2.78 $\pm$ 0.13 \\ 
88 & Gaia DR2 6454923636405184640 & 4835.79 $\pm$ 74.53 & 2.18 $\pm$ 0.17 &-0.49 $\pm$ 0.07 & 12.68 & 1.67 $\pm$ 0.05 & 2.77 $\pm$ 0.13 \\ 
89 & Gaia DR2 5791378917488589952 & 4730.86 $\pm$ 94.21 & 2.39 $\pm$ 0.18 &-0.03 $\pm$ 0.07 & 12.92 & 1.45 $\pm$ 0.05 & 2.77 $\pm$ 0.13 \\ 
90 & Gaia DR2 6108301045262477952 & 4842.90 $\pm$ 84.12 & 2.31 $\pm$ 0.19 &-0.07 $\pm$ 0.08 & 11.86 & 1.69 $\pm$ 0.07 & 2.77 $\pm$ 0.13 \\ 
91 & Gaia DR2 6129782620559721344 & 4774.26 $\pm$ 55.89 & 2.63 $\pm$ 0.16 & 0.14 $\pm$ 0.06 & 10.75 & 1.68 $\pm$ 0.02 & 2.77 $\pm$ 0.12 \\ 
92 & Gaia DR2 3200477756303359616 & 5154.93 $\pm$ 58.30 & 3.67 $\pm$ 0.17 & 0.06 $\pm$ 0.07 & 12.40 & 0.63 $\pm$ 0.02 & 2.77 $\pm$ 0.12 \\ 
93 & Gaia DR2 6145981622281078912 & 4886.66 $\pm$ 84.32 & 2.36 $\pm$ 0.17 &-0.16 $\pm$ 0.07 & 13.29 & 1.70 $\pm$ 0.10 & 2.76 $\pm$ 0.13 \\ 
94 & Gaia DR2 4200949362381698944 & 4809.22 $\pm$ 85.59 & 2.55 $\pm$ 0.19 & 0.04 $\pm$ 0.08 & 13.27 & 1.56 $\pm$ 0.11 & 2.76 $\pm$ 0.13 \\ 
95 & Gaia DR2 5702816893698573952 & 4975.65 $\pm$ 59.50 & 2.28 $\pm$ 0.15 &-0.52 $\pm$ 0.06 & 12.58 & 1.65 $\pm$ 0.06 & 2.75 $\pm$ 0.12 \\ 
96 & Gaia DR2 6719709224902437888 & 4873.48 $\pm$ 93.08 & 2.21 $\pm$ 0.17 &-0.46 $\pm$ 0.07 & 13.01 & 1.60 $\pm$ 0.09 & 2.74 $\pm$ 0.13 \\ 
97 & Gaia DR2 5431861567506547072 & 4933.04 $\pm$ 62.21 & 2.56 $\pm$ 0.16 & 0.06 $\pm$ 0.07 & 11.52 & 1.71 $\pm$ 0.05 & 2.74 $\pm$ 0.12 \\ 
98 & Gaia DR2 5773485774490988672 & 4902.15 $\pm$ 70.78 & 2.56 $\pm$ 0.16 &-0.14 $\pm$ 0.06 & 12.87 & 1.83 $\pm$ 0.06 & 2.73 $\pm$ 0.12 \\ 
99 & Gaia DR2 5373941081663359488 & 4638.57 $\pm$ 86.06 & 2.37 $\pm$ 0.18 &-0.04 $\pm$ 0.07 & 13.28 & 1.55 $\pm$ 0.05 & 2.73 $\pm$ 0.13 \\ 
100 & Gaia DR2 6779777645365765504 & 4723.99 $\pm$ 85.78 & 2.50 $\pm$ 0.16 & 0.14 $\pm$ 0.06 & 12.92 & 1.60 $\pm$ 0.08 & 2.72 $\pm$ 0.12 \\
101 & Gaia DR2 5452027607189177472 & 4655.45 $\pm$ 72.67 & 2.43 $\pm$ 0.18 &-0.32 $\pm$ 0.07 & 13.73 & 1.79 $\pm$ 0.08 & 2.71 $\pm$ 0.13 \\ 
102 & Gaia DR2 5675925248761091456 & 4851.50 $\pm$ 81.58 & 2.42 $\pm$ 0.18 &-0.09 $\pm$ 0.07 & 12.76 & 1.62 $\pm$ 0.06 & 2.71 $\pm$ 0.13 \\ 
103 & Gaia DR2 3073566459164534016 & 4649.46 $\pm$ 51.93 & 2.44 $\pm$ 0.14 &-0.01 $\pm$ 0.05 & 11.93 & 1.60 $\pm$ 0.05 & 2.69 $\pm$ 0.12 \\ 
104 & Gaia DR2 5817258053953332736 & 4720.67 $\pm$ 81.08 & 2.31 $\pm$ 0.19 &-0.05 $\pm$ 0.08 & 11.51 & 1.65 $\pm$ 0.04 & 2.68 $\pm$ 0.13 \\ 
105 & Gaia DR2 5391227736613349248 & 4779.03 $\pm$ 89.28 & 2.02 $\pm$ 0.18 &-0.71 $\pm$ 0.07 & 12.43 & 1.87 $\pm$ 0.06 & 2.67 $\pm$ 0.13 \\ 
106 & Gaia DR2 5584495698654750080 & 4619.57 $\pm$ 92.18 & 2.17 $\pm$ 0.16 &-0.44 $\pm$ 0.06 & 12.29 & 1.76 $\pm$ 0.05 & 2.65 $\pm$ 0.12 \\ 
107 & Gaia DR2 4235985476499510528 & 4435.76 $\pm$ 73.75 & 2.33 $\pm$ 0.19 &-0.14 $\pm$ 0.08 & 13.06 & 1.85 $\pm$ 0.09 & 2.64 $\pm$ 0.13 \\ 
108 & Gaia DR2 5808624207616553216 & 4800.47 $\pm$ 83.44 & 2.65 $\pm$ 0.20 & 0.19 $\pm$ 0.08 & 11.00 & 1.63 $\pm$ 0.03 & 2.64 $\pm$ 0.14 \\ 
109 & Gaia DR2 5791204782336493184 & 4391.82 $\pm$ 73.05 & 1.94 $\pm$ 0.19 &-0.10 $\pm$ 0.08 & 10.97 & 2.29 $\pm$ 0.04 & 2.62 $\pm$ 0.13 \\ 
110 & Gaia DR2 5811042927036279168 & 4740.72 $\pm$ 67.51 & 2.39 $\pm$ 0.17 &-0.18 $\pm$ 0.07 & 12.84 & 1.67 $\pm$ 0.07 & 2.60 $\pm$ 0.12 \\ 
111 & Gaia DR2 6654103157075218944 & 4898.66 $\pm$ 76.36 & 2.15 $\pm$ 0.18 &-0.48 $\pm$ 0.07 & 13.18 & 1.65 $\pm$ 0.07 & 2.59 $\pm$ 0.13 \\ 
112 & Gaia DR2 3466104544210064000 & 4838.46 $\pm$ 39.09 & 2.62 $\pm$ 0.12 & 0.04 $\pm$ 0.04 & 12.63 & 1.75 $\pm$ 0.06 & 2.59 $\pm$ 0.11 \\ 
113 & Gaia DR2 5624687732033481344 & 4842.71 $\pm$ 38.21 & 2.61 $\pm$ 0.11 & 0.15 $\pm$ 0.04 &  9.95 & 1.61 $\pm$ 0.02 & 2.59 $\pm$ 0.11 \\ 
114 & Gaia DR2 6779256987952631808 & 4827.94 $\pm$ 63.27 & 2.24 $\pm$ 0.16 &-0.53 $\pm$ 0.06 & 12.61 & 1.78 $\pm$ 0.09 & 2.58 $\pm$ 0.12 \\ 
115 & Gaia DR2 3470350896772650880 & 4829.62 $\pm$ 58.23 & 2.32 $\pm$ 0.16 &-0.05 $\pm$ 0.06 & 11.85 & 1.80 $\pm$ 0.06 & 2.58 $\pm$ 0.12 \\ 
116 & Gaia DR2 6091506902806438528 & 4928.63 $\pm$ 61.28 & 2.46 $\pm$ 0.16 &-0.26 $\pm$ 0.06 & 11.67 & 1.62 $\pm$ 0.08 & 2.58 $\pm$ 0.12 \\ 
117 & Gaia DR2 5602534118921439488 & 5028.84 $\pm$ 88.91 & 2.50 $\pm$ 0.20 &-0.18 $\pm$ 0.08 & 13.48 & 1.86 $\pm$ 0.07 & 2.57 $\pm$ 0.14 \\ 
118 & Gaia DR2 6036403911204069248 & 4748.28 $\pm$ 49.36 & 2.64 $\pm$ 0.14 & 0.14 $\pm$ 0.05 & 12.29 & 1.56 $\pm$ 0.05 & 2.56 $\pm$ 0.12 \\ 
119 & Gaia DR2 5373053294747925888 & 4229.96 $\pm$ 77.37 & 1.71 $\pm$ 0.20 &-0.21 $\pm$ 0.08 & 13.92 & 2.48 $\pm$ 0.11 & 2.55 $\pm$ 0.13 \\ 
120 & Gaia DR2 2696081783518779520 & 4620.45 $\pm$ 95.71 & 2.22 $\pm$ 0.20 &-0.21 $\pm$ 0.08 & 13.24 & 1.81 $\pm$ 0.12 & 2.55 $\pm$ 0.14 \\ 
\hline
\end{tabular}
\end{table*}

\begin{table*}
\begin{center}
\contcaption{} \end{center}
\begin{tabular}{cccccccccccccc}
\hline
\hline
S.No.&Object ID& Teff & log{\it g} & [Fe/H] & m$_v$ & L/L$_\odot$ & A(Li)\\
 & & (K)& (dex) & (dex)& (mag) & (dex) & (dex)\\
\hline
\hline
121 & Gaia DR2 5915006974134206080 & 4624.17 $\pm$ 68.64 & 2.20 $\pm$ 0.17 &-0.14 $\pm$ 0.07 & 12.87 & 1.64 $\pm$ 0.09 & 2.55 $\pm$ 0.13 \\ 
122 & Gaia DR2 5807050222360066304 & 4749.61 $\pm$ 96.47 & 2.30 $\pm$ 0.19 & 0.03 $\pm$ 0.08 & 13.47 & 1.64 $\pm$ 0.04 & 2.54 $\pm$ 0.13 \\ 
123 & Gaia DR2 6139263125199366400 & 5166.11 $\pm$ 80.37 & 2.40 $\pm$ 0.19 &-0.40 $\pm$ 0.08 & 13.32 & 1.55 $\pm$ 0.06 & 2.54 $\pm$ 0.13 \\ 
124 & Gaia DR2 5817665315632544384 & 5049.77 $\pm$ 94.53 & 2.57 $\pm$ 0.19 &-0.33 $\pm$ 0.08 & 11.81 & 1.78 $\pm$ 0.06 & 2.53 $\pm$ 0.13 \\ 
125 & Gaia DR2 6755428784277693568 & 4698.75 $\pm$ 66.16 & 2.59 $\pm$ 0.18 & 0.14 $\pm$ 0.08 & 11.84 & 1.58 $\pm$ 0.05 & 2.53 $\pm$ 0.13 \\ 
126 & Gaia DR2 3011575439538999296 & 4857.51 $\pm$ 90.29 & 2.32 $\pm$ 0.17 &-0.13 $\pm$ 0.07 & 12.86 & 1.35 $\pm$ 0.34 & 2.52 $\pm$ 0.13 \\ 
127 & Gaia DR2 5436690382057729408 & 4872.04 $\pm$ 83.56 & 2.42 $\pm$ 0.19 &-0.01 $\pm$ 0.08 & 13.46 & 1.71 $\pm$ 0.05 & 2.51 $\pm$ 0.13 \\ 
128 & Gaia DR2 5707005521902321536 & 4746.37 $\pm$ 68.27 & 2.60 $\pm$ 0.18 & 0.09 $\pm$ 0.07 & 13.08 & 1.77 $\pm$ 0.09 & 2.51 $\pm$ 0.13 \\ 
129 & Gaia DR2 6708545574104290688 & 4941.86 $\pm$ 78.34 & 2.32 $\pm$ 0.18 &-0.55 $\pm$ 0.07 & 13.55 & 1.71 $\pm$ 0.11 & 2.50 $\pm$ 0.13 \\ 
130 & Gaia DR2 6133545699095132032 & 4918.91 $\pm$ 81.14 & 2.54 $\pm$ 0.19 &-0.10 $\pm$ 0.08 & 13.38 & 1.78 $\pm$ 0.10 & 2.50 $\pm$ 0.13 \\ 
131 & Gaia DR2 6402728307604533632 & 4746.04 $\pm$ 75.30 & 2.49 $\pm$ 0.18 & 0.15 $\pm$ 0.08 & 13.03 & 1.60 $\pm$ 0.07 & 2.50 $\pm$ 0.13 \\ 
132 & Gaia DR2 4377588177421757440 & 4710.89 $\pm$ 76.75 & 2.22 $\pm$ 0.18 &-0.25 $\pm$ 0.08 & 13.49 & 1.40 $\pm$ 0.04 & 2.50 $\pm$ 0.13 \\ 
133 & Gaia DR2 6086711932594628352 & 4740.37 $\pm$ 55.21 & 2.78 $\pm$ 0.16 & 0.28 $\pm$ 0.06 & 12.33 & 1.54 $\pm$ 0.06 & 2.50 $\pm$ 0.12 \\ 
134 & Gaia DR2 4168620151412529536 & 4560.09 $\pm$ 69.08 & 2.75 $\pm$ 0.18 & 0.22 $\pm$ 0.07 & 13.01 & 1.35 $\pm$ 0.06 & 2.50 $\pm$ 0.13 \\ 
135 & Gaia DR2 6357806896762379904 & 4155.81 $\pm$ 69.59 & 1.49 $\pm$ 0.19 &-0.55 $\pm$ 0.08 & 12.35 & 2.60 $\pm$ 0.11 & 2.49 $\pm$ 0.13 \\ 
136 & Gaia DR2 6711980899405600896 & 4985.90 $\pm$ 89.22 & 2.77 $\pm$ 0.18 &-0.29 $\pm$ 0.07 & 11.79 & 1.59 $\pm$ 0.04 & 2.49 $\pm$ 0.13 \\ 
137 & Gaia DR2 5922073123050467200 & 4685.69 $\pm$ 54.79 & 2.01 $\pm$ 0.15 &-0.18 $\pm$ 0.06 & 10.15 & 2.24 $\pm$ 0.04 & 2.49 $\pm$ 0.12 \\ 
138 & Gaia DR2 6650476658491541376 & 4721.10 $\pm$ 43.90 & 2.48 $\pm$ 0.13 & 0.10 $\pm$ 0.05 & 12.14 & 1.86 $\pm$ 0.07 & 2.48 $\pm$ 0.11 \\ 
139 & Gaia DR2 2941562426030898816 & 4915.92 $\pm$ 56.97 & 2.80 $\pm$ 0.17 & 0.19 $\pm$ 0.07 & 12.09 & 1.65 $\pm$ 0.05 & 2.48 $\pm$ 0.12 \\ 
140 & Gaia DR2 5791281541991683200 & 4727.44 $\pm$ 78.48 & 2.49 $\pm$ 0.20 & 0.15 $\pm$ 0.08 & 13.78 & 1.51 $\pm$ 0.04 & 2.47 $\pm$ 0.13 \\ 
141 & Gaia DR2 5362422391691840768 & 4772.38 $\pm$ 61.90 & 2.61 $\pm$ 0.17 &-0.02 $\pm$ 0.07 & 13.01 & 1.66 $\pm$ 0.06 & 2.47 $\pm$ 0.13 \\ 
142 & Gaia DR2 5780764743325202048 & 4605.54 $\pm$ 62.68 & 2.19 $\pm$ 0.15 &-0.51 $\pm$ 0.06 & 12.76 & 1.94 $\pm$ 0.06 & 2.46 $\pm$ 0.12 \\ 
143 & Gaia DR2 4620358661206107776 & 4725.09 $\pm$ 71.84 & 2.53 $\pm$ 0.19 & 0.25 $\pm$ 0.08 & 12.92 & 1.48 $\pm$ 0.04 & 2.46 $\pm$ 0.13 \\ 
144 & Gaia DR2 5467980455756182656 & 4934.78 $\pm$ 85.21 & 2.32 $\pm$ 0.19 &-0.52 $\pm$ 0.08 & 13.68 & 1.81 $\pm$ 0.08 & 2.45 $\pm$ 0.13 \\ 
145 & Gaia DR2 5224684057518392960 & 4697.79 $\pm$ 56.45 & 2.22 $\pm$ 0.15 &-0.07 $\pm$ 0.06 & 12.38 & 2.03 $\pm$ 0.05 & 2.45 $\pm$ 0.12 \\ 
146 & Gaia DR2 6141920541724000000 & 4827.23 $\pm$ 66.59 & 2.43 $\pm$ 0.17 &-0.10 $\pm$ 0.07 & 13.03 & 1.63 $\pm$ 0.08 & 2.45 $\pm$ 0.13 \\ 
147 & Gaia DR2 5370859734689721600 & 4809.22 $\pm$ 64.08 & 2.55 $\pm$ 0.18 & 0.08 $\pm$ 0.07 & 12.61 & 1.72 $\pm$ 0.05 & 2.45 $\pm$ 0.13 \\ 
148 & Gaia DR2 2706959522635043968 & 4728.07 $\pm$ 58.88 & 2.46 $\pm$ 0.16 &-0.01 $\pm$ 0.06 & 12.31 & 1.80 $\pm$ 0.11 & 2.45 $\pm$ 0.12 \\ 
149 & Gaia DR2 5436262049261115008 & 4249.33 $\pm$ 31.71 & 1.52 $\pm$ 0.10 &-0.25 $\pm$ 0.03 &  9.75 & 2.60 $\pm$ 0.04 & 2.44 $\pm$ 0.11 \\ 
150 & Gaia DR2 4487925023966389376 & 4480.22 $\pm$ 65.82 & 2.48 $\pm$ 0.18 &-0.08 $\pm$ 0.08 & 13.46 & 1.77 $\pm$ 0.08 & 2.43 $\pm$ 0.13 \\
151 & Gaia DR2 6702531009288403712 & 4897.72 $\pm$ 68.30 & 2.38 $\pm$ 0.17 &-0.21 $\pm$ 0.07 & 11.73 & 1.54 $\pm$ 0.05 & 2.43 $\pm$ 0.13 \\ 
152 & Gaia DR2 5914068472234966016 & 4717.91 $\pm$ 86.67 & 2.75 $\pm$ 0.19 & 0.13 $\pm$ 0.08 & 13.71 & 1.59 $\pm$ 0.06 & 2.42 $\pm$ 0.13 \\ 
153 & Gaia DR2 5421023814530120704 & 4778.86 $\pm$ 66.32 & 2.67 $\pm$ 0.18 & 0.20 $\pm$ 0.07 & 12.69 & 1.55 $\pm$ 0.05 & 2.42 $\pm$ 0.13 \\ 
154 & Gaia DR2 6635029069878586112 & 4441.37 $\pm$ 49.88 & 1.97 $\pm$ 0.15 &-0.37 $\pm$ 0.06 & 10.63 & 1.91 $\pm$ 0.04 & 2.42 $\pm$ 0.12 \\ 
155 & Gaia DR2 6157593667660508800 & 4743.33 $\pm$ 77.33 & 2.20 $\pm$ 0.19 &-0.32 $\pm$ 0.08 & 13.80 & 1.65 $\pm$ 0.12 & 2.41 $\pm$ 0.13 \\ 
156 & Gaia DR2 5792476539327714816 & 4690.75 $\pm$ 76.60 & 2.22 $\pm$ 0.18 &-0.35 $\pm$ 0.07 & 13.53 & 1.68 $\pm$ 0.04 & 2.41 $\pm$ 0.13 \\ 
157 & Gaia DR2 5633145995807169664 & 4755.24 $\pm$ 57.81 & 2.57 $\pm$ 0.17 & 0.15 $\pm$ 0.07 & 12.44 & 1.65 $\pm$ 0.05 & 2.41 $\pm$ 0.12 \\ 
158 & Gaia DR2 5376150928233716992 & 4625.72 $\pm$ 70.57 & 2.42 $\pm$ 0.18 & 0.03 $\pm$ 0.07 & 13.05 & 1.68 $\pm$ 0.06 & 2.40 $\pm$ 0.13 \\ 
159 & Gaia DR2 5909836829933180672 & 4729.30 $\pm$ 56.53 & 2.55 $\pm$ 0.16 & 0.10 $\pm$ 0.06 & 12.53 & 1.57 $\pm$ 0.05 & 2.40 $\pm$ 0.12 \\ 
160 & Gaia DR2 5297697956095835264 & 4769.33 $\pm$ 46.76 & 2.31 $\pm$ 0.14 &-0.13 $\pm$ 0.05 & 10.57 & 1.98 $\pm$ 0.03 & 2.40 $\pm$ 0.12 \\ 
161 & Gaia DR2 6146039827677625856 & 4770.25 $\pm$ 69.99 & 2.42 $\pm$ 0.19 &-0.01 $\pm$ 0.08 & 13.28 & 1.55 $\pm$ 0.12 & 2.39 $\pm$ 0.13 \\ 
162 & Gaia DR2 4047063815234837504 & 4798.05 $\pm$ 25.22 & 2.52 $\pm$ 0.08 & 0.05 $\pm$ 0.03 &  9.51 & 1.62 $\pm$ 0.02 & 2.38 $\pm$ 0.11 \\ 
163 & Gaia DR2 5493998989680647168 & 4905.04 $\pm$ 93.17 & 2.54 $\pm$ 0.19 &-0.08 $\pm$ 0.08 & 13.77 & 1.74 $\pm$ 0.05 & 2.37 $\pm$ 0.13 \\ 
164 & Gaia DR2 5793721766308320128 & 4289.69 $\pm$ 96.75 & 1.37 $\pm$ 0.19 &-0.70 $\pm$ 0.08 & 12.31 & 2.19 $\pm$ 0.07 & 2.36 $\pm$ 0.13 \\ 
165 & Gaia DR2 3459399967119296128 & 5054.88 $\pm$ 98.69 & 2.93 $\pm$ 0.19 &-0.26 $\pm$ 0.08 & 13.82 & 1.48 $\pm$ 0.06 & 2.35 $\pm$ 0.13 \\ 
166 & Gaia DR2 5807253009230496256 & 4757.73 $\pm$ 70.24 & 2.30 $\pm$ 0.18 &-0.06 $\pm$ 0.07 & 12.60 & 1.80 $\pm$ 0.05 & 2.35 $\pm$ 0.13 \\ 
167 & Gaia DR2 5721819516941191296 & 4778.66 $\pm$ 64.33 & 2.37 $\pm$ 0.16 &-0.20 $\pm$ 0.07 & 12.85 & 1.67 $\pm$ 0.06 & 2.35 $\pm$ 0.12 \\
168 & Gaia DR2 4471569208692643200 & 4666.33 $\pm$ 59.43 & 2.93 $\pm$ 0.17 & 0.38 $\pm$ 0.07 & 12.44 & 1.42 $\pm$ 0.04 & 2.35 $\pm$ 0.13 \\ 
169 & Gaia DR2 6082568869702190720 & 4874.32 $\pm$ 54.17 & 2.29 $\pm$ 0.15 &-0.10 $\pm$ 0.06 & 10.30 & 1.86 $\pm$ 0.04 & 2.35 $\pm$ 0.12 \\ 
170 & Gaia DR2 5791416507035670400 & 4881.80 $\pm$ 77.93 & 2.61 $\pm$ 0.16 &-0.12 $\pm$ 0.06 & 10.44 & 1.59 $\pm$ 0.03 & 2.35 $\pm$ 0.12 \\ 
171 & Gaia DR2 4037523623386846336 & 4775.57 $\pm$ 60.72 & 2.78 $\pm$ 0.18 & 0.19 $\pm$ 0.07 & 11.47 & 1.56 $\pm$ 0.04 & 2.34 $\pm$ 0.13 \\ 
172 & Gaia DR2 6637733314425753984 & 4714.68 $\pm$ 54.08 & 2.17 $\pm$ 0.15 &-0.28 $\pm$ 0.06 & 12.51 & 1.63 $\pm$ 0.06 & 2.33 $\pm$ 0.12 \\ 
173 & Gaia DR2 5374556498937060608 & 4978.66 $\pm$ 57.47 & 2.49 $\pm$ 0.16 &-0.16 $\pm$ 0.06 & 11.53 & 1.72 $\pm$ 0.04 & 2.33 $\pm$ 0.12 \\ 
174 & Gaia DR2 5818798160506192000 & 4763.25 $\pm$ 65.20 & 2.52 $\pm$ 0.18 & 0.16 $\pm$ 0.07 & 11.84 & 1.59 $\pm$ 0.03 & 2.33 $\pm$ 0.13 \\ 
175 & Gaia DR2 5820639666385663360 & 4751.42 $\pm$ 69.01 & 2.43 $\pm$ 0.18 & 0.01 $\pm$ 0.07 & 12.79 & 1.53 $\pm$ 0.05 & 2.32 $\pm$ 0.13 \\ 
176 & Gaia DR2 5914162617918721280 & 4737.86 $\pm$ 55.13 & 2.56 $\pm$ 0.16 & 0.22 $\pm$ 0.06 & 12.20 & 1.56 $\pm$ 0.05 & 2.32 $\pm$ 0.12 \\ 
177 & Gaia DR2 6070503756813117056 & 4901.75 $\pm$ 56.54 & 2.58 $\pm$ 0.16 & 0.07 $\pm$ 0.06 & 11.12 & 1.75 $\pm$ 0.05 & 2.32 $\pm$ 0.12 \\ 
178 & Gaia DR2 5398253585851636992 & 4600.65 $\pm$ 73.73 & 2.37 $\pm$ 0.19 & 0.40 $\pm$ 0.08 & 13.62 & 1.60 $\pm$ 0.05 & 2.31 $\pm$ 0.13 \\ 
179 & Gaia DR2 5386860304630456320 & 5109.17 $\pm$ 71.30 & 2.50 $\pm$ 0.18 &-0.16 $\pm$ 0.07 & 12.29 & 1.57 $\pm$ 0.04 & 2.31 $\pm$ 0.13 \\ 
180 & Gaia DR2 3253963778612870912 & 4758.28 $\pm$ 52.62 & 2.45 $\pm$ 0.15 & 0.07 $\pm$ 0.06 & 12.20 & 1.63 $\pm$ 0.05 & 2.31 $\pm$ 0.12 \\ 
\hline
\end{tabular}
\end{table*}

\begin{table*}
\begin{center}
\contcaption{} \end{center}
\begin{tabular}{cccccccccccccc}
\hline
\hline
S.No.&Object ID& Teff & log{\it g} & [Fe/H] & m$_v$ & L/L$_\odot$ & A(Li)\\
 & & (K)& (dex) & (dex)& (mag) & (dex) & (dex)\\
\hline
\hline
181 & Gaia DR2 5229285758501546624 & 4750.31 $\pm$ 56.69 & 2.65 $\pm$ 0.16 & 0.12 $\pm$ 0.06 & 11.59 & 1.54 $\pm$ 0.02 & 2.31 $\pm$ 0.12 \\ 
182 & Gaia DR2 2895200315655057920 & 4848.37 $\pm$ 78.74 & 2.32 $\pm$ 0.18 &-0.25 $\pm$ 0.08 & 13.84 & 1.87 $\pm$ 0.08 & 2.29 $\pm$ 0.13 \\ 
183 & Gaia DR2 5416314370010160512 & 4832.31 $\pm$ 62.17 & 2.43 $\pm$ 0.17 & 0.06 $\pm$ 0.07 & 13.15 & 1.82 $\pm$ 0.09 & 2.29 $\pm$ 0.13 \\ 
184 & Gaia DR2 6078451610978783488 & 4886.28 $\pm$ 64.56 & 2.62 $\pm$ 0.18 & 0.19 $\pm$ 0.07 & 12.01 & 1.72 $\pm$ 0.05 & 2.29 $\pm$ 0.13 \\ 
185 & Gaia DR2 5363510495882765952 & 4767.83 $\pm$ 65.11 & 2.66 $\pm$ 0.18 & 0.18 $\pm$ 0.07 & 11.52 & 1.61 $\pm$ 0.04 & 2.29 $\pm$ 0.13 \\ 
186 & Gaia DR2 6715331042616800768 & 4773.35 $\pm$ 58.89 & 2.42 $\pm$ 0.16 &-0.06 $\pm$ 0.07 & 10.89 & 1.60 $\pm$ 0.05 & 2.29 $\pm$ 0.12 \\ 
187 & Gaia DR2 3463132117603485440 & 4776.95 $\pm$ 60.66 & 2.65 $\pm$ 0.17 & 0.12 $\pm$ 0.07 & 12.44 & 1.67 $\pm$ 0.06 & 2.28 $\pm$ 0.13 \\ 
188 & Gaia DR2 6702221629897848960 & 4782.02 $\pm$ 64.82 & 2.73 $\pm$ 0.18 & 0.26 $\pm$ 0.07 & 11.69 & 1.58 $\pm$ 0.04 & 2.28 $\pm$ 0.13 \\ 
189 & Gaia DR2 5374651125652223616 & 4926.98 $\pm$ 67.21 & 2.43 $\pm$ 0.13 &-0.23 $\pm$ 0.05 &  9.69 & 1.72 $\pm$ 0.02 & 2.28 $\pm$ 0.11 \\ 
190 & Gaia DR2 5583693368703146496 & 4676.01 $\pm$ 82.42 & 2.53 $\pm$ 0.20 & 0.18 $\pm$ 0.09 & 13.62 & 1.61 $\pm$ 0.04 & 2.26 $\pm$ 0.14 \\ 
191 & Gaia DR2 6090505587008797440 & 4529.86 $\pm$ 56.53 & 2.40 $\pm$ 0.16 &-0.16 $\pm$ 0.06 & 12.08 & 1.63 $\pm$ 0.12 & 2.26 $\pm$ 0.12 \\ 
192 & Gaia DR2 5907299672495709568 & 4556.19 $\pm$ 77.41 & 2.60 $\pm$ 0.20 & 0.04 $\pm$ 0.08 & 13.23 & 1.72 $\pm$ 0.08 & 2.25 $\pm$ 0.14 \\ 
193 & Gaia DR2 5797558825674487936 & 4752.38 $\pm$ 67.70 & 2.32 $\pm$ 0.15 &-0.09 $\pm$ 0.06 & 13.29 & 1.63 $\pm$ 0.05 & 2.25 $\pm$ 0.12 \\ 
194 & Gaia DR2 5247038610680907904 & 4970.98 $\pm$ 58.13 & 2.53 $\pm$ 0.17 &-0.07 $\pm$ 0.07 & 11.80 & 1.84 $\pm$ 0.03 & 2.25 $\pm$ 0.12 \\ 
195 & Gaia DR2 6721954290205665920 & 4721.79 $\pm$ 63.27 & 2.64 $\pm$ 0.18 & 0.17 $\pm$ 0.07 & 11.27 & 1.60 $\pm$ 0.03 & 2.24 $\pm$ 0.13 \\ 
196 & Gaia DR2 6703068017635031296 & 5057.53 $\pm$ 38.79 & 2.67 $\pm$ 0.12 & 0.14 $\pm$ 0.05 &  9.92 & 1.82 $\pm$ 0.03 & 2.24 $\pm$ 0.11 \\ 
197 & Gaia DR2 2938523856928065536 & 5116.55 $\pm$ 81.32 & 2.53 $\pm$ 0.20 &-0.22 $\pm$ 0.08 & 13.72 & 1.87 $\pm$ 0.08 & 2.22 $\pm$ 0.14 \\ 
198 & Gaia DR2 6092109263376811648 & 4700.12 $\pm$ 69.35 & 2.88 $\pm$ 0.19 & 0.41 $\pm$ 0.08 & 14.00 & 1.43 $\pm$ 0.07 & 2.22 $\pm$ 0.13 \\ 
199 & Gaia DR2 5376045542618548736 & 4760.51 $\pm$ 57.42 & 2.47 $\pm$ 0.16 & 0.03 $\pm$ 0.06 & 12.56 & 1.76 $\pm$ 0.07 & 2.22 $\pm$ 0.12 \\ 
200 & Gaia DR2 5420436095505648128 & 4924.01 $\pm$ 59.30 & 2.55 $\pm$ 0.16 & 0.05 $\pm$ 0.06 & 12.21 & 1.69 $\pm$ 0.05 & 2.22 $\pm$ 0.12 \\
201 & Gaia DR2 5315969124930734464 & 4353.50 $\pm$ 30.73 & 1.69 $\pm$ 0.10 &-0.03 $\pm$ 0.04 &  9.08 & 2.50 $\pm$ 0.02 & 2.22 $\pm$ 0.11 \\ 
202 & Gaia DR2 3080469021370104192 & 4962.88 $\pm$ 50.63 & 2.64 $\pm$ 0.15 & 0.10 $\pm$ 0.06 &  9.57 & 1.82 $\pm$ 0.02 & 2.21 $\pm$ 0.12 \\ 
203 & Gaia DR2 5415764025783016832 & 4782.19 $\pm$ 59.20 & 2.62 $\pm$ 0.17 & 0.21 $\pm$ 0.07 & 11.86 & 1.46 $\pm$ 0.03 & 2.20 $\pm$ 0.13 \\ 
204 & Gaia DR2 4359779314569319296 & 4797.57 $\pm$ 51.89 & 2.61 $\pm$ 0.15 & 0.10 $\pm$ 0.06 & 12.55 & 1.23 $\pm$ 0.05 & 2.20 $\pm$ 0.12 \\ 
205 & Gaia DR2 5793765785434358528 & 4805.80 $\pm$ 56.98 & 2.57 $\pm$ 0.17 & 0.22 $\pm$ 0.07 & 10.33 & 1.79 $\pm$ 0.02 & 2.19 $\pm$ 0.12 \\ 
206 & Gaia DR2 5455533812332089728 & 4729.53 $\pm$ 70.73 & 2.37 $\pm$ 0.19 & 0.09 $\pm$ 0.08 & 13.46 & 1.73 $\pm$ 0.05 & 2.18 $\pm$ 0.13 \\ 
207 & Gaia DR2 6735120254754428160 & 4830.97 $\pm$ 71.85 & 2.44 $\pm$ 0.19 & 0.12 $\pm$ 0.08 & 11.96 & 1.63 $\pm$ 0.04 & 2.18 $\pm$ 0.13 \\ 
208 & Gaia DR2 5822073571279630208 & 4857.42 $\pm$ 58.49 & 2.61 $\pm$ 0.17 & 0.09 $\pm$ 0.07 & 11.26 & 1.46 $\pm$ 0.03 & 2.17 $\pm$ 0.12 \\ 
209 & Gaia DR2 5678212366090166528 & 4863.77 $\pm$ 96.11 & 2.21 $\pm$ 0.20 &-0.20 $\pm$ 0.08 & 13.47 & 1.96 $\pm$ 0.10 & 2.16 $\pm$ 0.13 \\ 
210 & Gaia DR2 5574618137164897536 & 4778.78 $\pm$ 71.35 & 2.26 $\pm$ 0.19 &-0.19 $\pm$ 0.08 & 13.67 & 1.64 $\pm$ 0.04 & 2.16 $\pm$ 0.13 \\ 
211 & Gaia DR2 5384495873591911680 & 4202.23 $\pm$ 55.36 & 1.97 $\pm$ 0.16 &-0.42 $\pm$ 0.06 & 12.51 & 2.16 $\pm$ 0.08 & 2.16 $\pm$ 0.12 \\ 
212 & Gaia DR2 6081398881948534144 & 4936.66 $\pm$ 49.00 & 2.69 $\pm$ 0.15 & 0.04 $\pm$ 0.06 & 12.36 & 1.67 $\pm$ 0.07 & 2.16 $\pm$ 0.12 \\ 
213 & Gaia DR2 5807170202261464704 & 4965.75 $\pm$ 78.40 & 2.26 $\pm$ 0.18 &-0.28 $\pm$ 0.07 & 12.10 & 1.62 $\pm$ 0.05 & 2.16 $\pm$ 0.13 \\ 
214 & Gaia DR2 6142927861878879104 & 4672.71 $\pm$ 74.63 & 2.36 $\pm$ 0.17 &-0.03 $\pm$ 0.07 & 11.68 & 1.64 $\pm$ 0.04 & 2.16 $\pm$ 0.12 \\ 
215 & Gaia DR2 5947407966903218432 & 4932.36 $\pm$ 84.69 & 2.63 $\pm$ 0.21 & 0.09 $\pm$ 0.09 & 11.17 & 1.72 $\pm$ 0.05 & 2.16 $\pm$ 0.14 \\ 
216 & Gaia DR2 6002549815052973568 & 4810.41 $\pm$ 70.69 & 2.41 $\pm$ 0.18 &-0.05 $\pm$ 0.07 & 13.37 & 1.71 $\pm$ 0.08 & 2.15 $\pm$ 0.13 \\ 
217 & Gaia DR2 4038214773873148032 & 4934.93 $\pm$ 97.20 & 2.31 $\pm$ 0.19 &-0.26 $\pm$ 0.08 & 13.09 & 1.73 $\pm$ 0.10 & 2.15 $\pm$ 0.13 \\ 
218 & Gaia DR2 3142806554658429952 & 4692.94 $\pm$ 85.60 & 2.06 $\pm$ 0.16 &-0.50 $\pm$ 0.07 & 10.03 & 2.17 $\pm$ 0.04 & 2.15 $\pm$ 0.12 \\ 
219 & Gaia DR2 6703670653085991296 & 4619.50 $\pm$ 81.71 & 2.18 $\pm$ 0.19 & 0.14 $\pm$ 0.08 & 13.47 & 1.59 $\pm$ 0.07 & 2.14 $\pm$ 0.13 \\ 
220 & Gaia DR2 6788265634612883328 & 4743.15 $\pm$ 64.51 & 2.35 $\pm$ 0.17 &-0.23 $\pm$ 0.07 & 13.36 & 1.74 $\pm$ 0.11 & 2.13 $\pm$ 0.13 \\ 
221 & Gaia DR2 6093100919786226944 & 4526.00 $\pm$ 89.56 & 2.25 $\pm$ 0.19 & 0.00 $\pm$ 0.08 & 12.82 & 1.63 $\pm$ 0.07 & 2.13 $\pm$ 0.13 \\ 
222 & Gaia DR2 2920138270166122496 & 4714.61 $\pm$ 54.33 & 2.67 $\pm$ 0.16 & 0.13 $\pm$ 0.06 & 11.45 & 1.67 $\pm$ 0.03 & 2.13 $\pm$ 0.12 \\ 
223 & Gaia DR2 5793537602412430208 & 4805.38 $\pm$ 75.45 & 2.21 $\pm$ 0.18 &-0.20 $\pm$ 0.07 & 11.11 & 1.69 $\pm$ 0.02 & 2.13 $\pm$ 0.13 \\ 
224 & Gaia DR2 5918644566604591360 & 4920.95 $\pm$ 75.87 & 2.36 $\pm$ 0.17 &-0.20 $\pm$ 0.07 & 12.88 & 1.86 $\pm$ 0.10 & 2.12 $\pm$ 0.12 \\ 
225 & Gaia DR2 5387273931457236096 & 4702.64 $\pm$ 96.11 & 2.46 $\pm$ 0.20 & 0.01 $\pm$ 0.09 & 13.28 & 1.63 $\pm$ 0.06 & 2.12 $\pm$ 0.14 \\ 
226 & Gaia DR2 5492185242170706304 & 4784.87 $\pm$ 68.07 & 2.31 $\pm$ 0.18 &-0.14 $\pm$ 0.07 & 13.77 & 1.75 $\pm$ 0.05 & 2.11 $\pm$ 0.13 \\ 
227 & Gaia DR2 5806980987485756288 & 4985.32 $\pm$ 88.98 & 2.55 $\pm$ 0.19 &-0.01 $\pm$ 0.08 & 13.98 & 1.54 $\pm$ 0.06 & 2.11 $\pm$ 0.13 \\ 
228 & Gaia DR2 5707045447918185728 & 4837.75 $\pm$ 60.39 & 2.30 $\pm$ 0.17 &-0.13 $\pm$ 0.07 & 11.50 & 1.80 $\pm$ 0.06 & 2.11 $\pm$ 0.12 \\ 
229 & Gaia DR2 5402146368112930176 & 4753.24 $\pm$ 69.36 & 2.43 $\pm$ 0.19 & 0.04 $\pm$ 0.08 & 13.85 & 1.70 $\pm$ 0.09 & 2.09 $\pm$ 0.13 \\ 
230 & Gaia DR2 5463505439134617600 & 4621.60 $\pm$ 52.92 & 2.48 $\pm$ 0.16 & 0.05 $\pm$ 0.06 & 12.32 & 1.59 $\pm$ 0.05 & 2.09 $\pm$ 0.12 \\ 
231 & Gaia DR2 5420381669680279168 & 4838.36 $\pm$ 56.72 & 2.70 $\pm$ 0.17 & 0.21 $\pm$ 0.07 & 11.73 & 1.71 $\pm$ 0.04 & 2.09 $\pm$ 0.12 \\ 
232 & Gaia DR2 5947033067785490304 & 4555.43 $\pm$ 88.81 & 2.41 $\pm$ 0.19 & 0.11 $\pm$ 0.08 & 10.74 & 1.78 $\pm$ 0.04 & 2.09 $\pm$ 0.13 \\ 
233 & Gaia DR2 6007329426452855296 & 4648.61 $\pm$ 99.05 & 2.57 $\pm$ 0.20 & 0.14 $\pm$ 0.08 & 13.43 & 1.77 $\pm$ 0.11 & 2.08 $\pm$ 0.13 \\ 
234 & Gaia DR2 5913247721159217152 & 4897.77 $\pm$ 66.21 & 2.57 $\pm$ 0.18 & 0.07 $\pm$ 0.08 & 11.96 & 1.71 $\pm$ 0.04 & 2.08 $\pm$ 0.13 \\ 
235 & Gaia DR2 5947058837590423552 & 4653.73 $\pm$ 94.84 & 2.47 $\pm$ 0.19 &-0.01 $\pm$ 0.08 & 11.36 & 1.46 $\pm$ 0.04 & 2.08 $\pm$ 0.13 \\ 
236 & Gaia DR2 5791153204062707456 & 4851.14 $\pm$ 68.96 & 2.46 $\pm$ 0.17 &-0.07 $\pm$ 0.07 & 12.82 & 1.74 $\pm$ 0.05 & 2.07 $\pm$ 0.13 \\ 
237 & Gaia DR2 6041508604395805568 & 4676.47 $\pm$ 58.77 & 2.62 $\pm$ 0.17 & 0.17 $\pm$ 0.07 & 12.94 & 1.71 $\pm$ 0.09 & 2.07 $\pm$ 0.13 \\ 
238 & Gaia DR2 5778543321820636928 & 4884.46 $\pm$ 55.04 & 2.25 $\pm$ 0.15 &-0.36 $\pm$ 0.06 & 12.51 & 1.55 $\pm$ 0.04 & 2.07 $\pm$ 0.12 \\ 
239 & Gaia DR2 6635692968739915264 & 4784.09 $\pm$ 89.73 & 2.29 $\pm$ 0.20 &-0.27 $\pm$ 0.08 & 13.22 & 1.66 $\pm$ 0.12 & 2.06 $\pm$ 0.14 \\ 
240 & Gaia DR2 6458252540999316864 & 4781.76 $\pm$ 76.96 & 2.46 $\pm$ 0.18 &-0.25 $\pm$ 0.07 & 12.99 & 1.84 $\pm$ 0.09 & 2.05 $\pm$ 0.13 \\ 
\hline
\end{tabular}
\end{table*}

\begin{table*}
\begin{center}
\contcaption{} \end{center}
\begin{tabular}{cccccccccccccc}
\hline
\hline
S.No.&Object ID& Teff & log{\it g} & [Fe/H] & m$_v$ & L/L$_\odot$ & A(Li)\\
 & & (K)& (dex) & (dex)& (mag) & (dex) & (dex)\\
\hline
\hline
241 & Gaia DR2 5780959051946613888 & 4730.17 $\pm$ 89.76 & 2.23 $\pm$ 0.20 &-0.16 $\pm$ 0.08 & 12.22 & 2.08 $\pm$ 0.07 & 2.05 $\pm$ 0.13 \\ 
242 & Gaia DR2 6099195650177688960 & 4767.01 $\pm$ 78.98 & 2.51 $\pm$ 0.17 & 0.15 $\pm$ 0.07 & 13.32 & 1.64 $\pm$ 0.10 & 2.05 $\pm$ 0.13 \\ 
243 & Gaia DR2 5946690054532114688 & 4789.73 $\pm$ 58.84 & 2.89 $\pm$ 0.17 & 0.32 $\pm$ 0.07 & 13.50 & 1.46 $\pm$ 0.04 & 2.05 $\pm$ 0.13 \\ 
244 & Gaia DR2 1753307668590290304 & 4816.85 $\pm$ 43.58 & 2.38 $\pm$ 0.13 & 0.02 $\pm$ 0.05 &  9.99 & 1.77 $\pm$ 0.02 & 2.05 $\pm$ 0.12 \\ 
245 & Gaia DR2 6141356694120988416 & 4732.71 $\pm$ 75.38 & 2.59 $\pm$ 0.20 & 0.08 $\pm$ 0.08 & 12.76 & 1.65 $\pm$ 0.06 & 2.04 $\pm$ 0.13 \\ 
246 & Gaia DR2 5432224402043330688 & 4421.82 $\pm$ 57.08 & 2.16 $\pm$ 0.17 &-0.13 $\pm$ 0.07 & 11.53 & 2.08 $\pm$ 0.05 & 2.04 $\pm$ 0.12 \\ 
247 & Gaia DR2 6635938262913646848 & 4752.53 $\pm$ 89.22 & 2.25 $\pm$ 0.20 &-0.14 $\pm$ 0.09 & 12.42 & 1.52 $\pm$ 0.05 & 2.04 $\pm$ 0.14 \\ 
248 & Gaia DR2 2955975717800052352 & 4952.24 $\pm$ 63.53 & 2.28 $\pm$ 0.17 &-0.47 $\pm$ 0.07 & 12.85 & 1.76 $\pm$ 0.05 & 2.03 $\pm$ 0.13 \\ 
249 & Gaia DR2 4141066630537415552 & 4654.90 $\pm$ 58.34 & 3.01 $\pm$ 0.17 & 0.53 $\pm$ 0.07 & 12.98 & 1.13 $\pm$ 0.04 & 2.03 $\pm$ 0.12 \\ 
250 & Gaia DR2 6092954547301602944 & 4709.12 $\pm$ 88.39 & 2.36 $\pm$ 0.20 &-0.03 $\pm$ 0.08 & 14.04 & 1.70 $\pm$ 0.12 & 2.02 $\pm$ 0.13 \\
251 & Gaia DR2 6005001897776085248 & 4806.41 $\pm$ 52.60 & 2.52 $\pm$ 0.16 &-0.04 $\pm$ 0.06 & 13.06 & 1.68 $\pm$ 0.09 & 2.02 $\pm$ 0.12 \\ 
252 & Gaia DR2 5820236081207169024 & 4608.44 $\pm$ 80.85 & 2.25 $\pm$ 0.20 &-0.13 $\pm$ 0.08 & 11.56 & 1.90 $\pm$ 0.04 & 2.02 $\pm$ 0.13 \\ 
253 & Gaia DR2 5437171658912631040 & 4657.19 $\pm$ 65.97 & 2.76 $\pm$ 0.19 & 0.25 $\pm$ 0.08 & 12.48 & 1.54 $\pm$ 0.04 & 2.02 $\pm$ 0.13 \\ 
254 & Gaia DR2 6128997431812095360 & 4757.61 $\pm$ 64.48 & 2.59 $\pm$ 0.18 & 0.14 $\pm$ 0.07 & 13.62 & 1.69 $\pm$ 0.10 & 2.01 $\pm$ 0.13 \\ 
255 & Gaia DR2 5647746204556453248 & 4822.54 $\pm$ 70.92 & 2.46 $\pm$ 0.18 &-0.04 $\pm$ 0.08 & 13.81 & 1.55 $\pm$ 0.07 & 2.01 $\pm$ 0.13 \\ 
256 & Gaia DR2 5455081328937070720 & 4679.82 $\pm$ 65.84 & 2.69 $\pm$ 0.18 & 0.17 $\pm$ 0.08 & 13.28 & 1.66 $\pm$ 0.07 & 2.01 $\pm$ 0.13 \\ 
257 & Gaia DR2 4471847934887193216 & 4598.36 $\pm$ 83.97 & 2.22 $\pm$ 0.19 & 0.18 $\pm$ 0.08 & 13.47 & 1.58 $\pm$ 0.06 & 2.01 $\pm$ 0.13 \\ 
258 & Gaia DR2 4139228792550926976 & 4620.55 $\pm$ 66.06 & 3.04 $\pm$ 0.19 & 0.56 $\pm$ 0.08 & 13.68 & 1.22 $\pm$ 0.07 & 2.01 $\pm$ 0.13 \\ 
259 & Gaia DR2 5822454964399685376 & 4859.07 $\pm$ 65.27 & 2.60 $\pm$ 0.18 & 0.01 $\pm$ 0.07 & 11.77 & 1.70 $\pm$ 0.04 & 2.01 $\pm$ 0.13 \\ 
260 & Gaia DR2 5703082464414878464 & 4701.55 $\pm$ 88.21 & 2.12 $\pm$ 0.19 &-0.28 $\pm$ 0.08 & 13.47 & 1.84 $\pm$ 0.06 & 2.00 $\pm$ 0.13 \\ 
261 & Gaia DR2 5916687886896990592 & 4722.45 $\pm$ 86.31 & 2.59 $\pm$ 0.19 & 0.20 $\pm$ 0.08 & 13.40 & 1.72 $\pm$ 0.06 & 2.00 $\pm$ 0.13 \\ 
262 & Gaia DR2 6392623452147567744 & 4761.13 $\pm$ 38.41 & 2.34 $\pm$ 0.12 &-0.13 $\pm$ 0.04 & 12.06 & 1.90 $\pm$ 0.04 & 2.00 $\pm$ 0.11 \\ 
263 & Gaia DR2 6009730721185313280 & 4081.66 $\pm$ 55.72 & 1.55 $\pm$ 0.16 &-0.54 $\pm$ 0.06 & 12.51 & 2.42 $\pm$ 0.10 & 1.99 $\pm$ 0.12 \\ 
264 & Gaia DR2 6385288266481482880 & 4995.72 $\pm$ 93.12 & 2.33 $\pm$ 0.20 &-0.48 $\pm$ 0.08 & 13.63 & 1.78 $\pm$ 0.05 & 1.98 $\pm$ 0.13 \\ 
265 & Gaia DR2 6852586167190048768 & 4862.05 $\pm$ 73.48 & 2.26 $\pm$ 0.19 &-0.22 $\pm$ 0.08 & 12.35 & 1.40 $\pm$ 0.06 & 1.98 $\pm$ 0.13 \\ 
266 & Gaia DR2 6386541086965979904 & 4870.52 $\pm$ 58.50 & 2.03 $\pm$ 0.15 &-0.44 $\pm$ 0.06 & 12.57 & 2.25 $\pm$ 0.08 & 1.96 $\pm$ 0.12 \\ 
267 & Gaia DR2 5561870051457417600 & 4761.80 $\pm$ 66.12 & 2.49 $\pm$ 0.18 &-0.05 $\pm$ 0.07 & 13.60 & 1.60 $\pm$ 0.04 & 1.96 $\pm$ 0.13 \\ 
268 & Gaia DR2 5199361067780431616 & 4505.28 $\pm$ 77.93 & 2.48 $\pm$ 0.19 &-0.04 $\pm$ 0.08 & 12.87 & 1.66 $\pm$ 0.04 & 1.96 $\pm$ 0.13 \\ 
269 & Gaia DR2 5295513776243667328 & 4540.07 $\pm$ 76.27 & 2.43 $\pm$ 0.17 & 0.09 $\pm$ 0.07 & 13.00 & 1.53 $\pm$ 0.03 & 1.96 $\pm$ 0.13 \\ 
270 & Gaia DR2 6037582420180899328 & 4803.14 $\pm$ 53.99 & 2.53 $\pm$ 0.16 & 0.00 $\pm$ 0.06 & 12.46 & 1.60 $\pm$ 0.06 & 1.96 $\pm$ 0.12 \\ 
271 & Gaia DR2 5296323669638828416 & 4517.15 $\pm$ 92.60 & 2.31 $\pm$ 0.20 &-0.02 $\pm$ 0.08 & 13.60 & 1.55 $\pm$ 0.03 & 1.95 $\pm$ 0.13 \\ 
272 & Gaia DR2 6711971725355453184 & 4863.09 $\pm$ 47.99 & 2.34 $\pm$ 0.14 &-0.08 $\pm$ 0.05 & 10.55 & 1.99 $\pm$ 0.04 & 1.95 $\pm$ 0.12 \\ 
273 & Gaia DR2 5296334596035138048 & 5170.73 $\pm$ 64.03 & 3.91 $\pm$ 0.19 & 0.34 $\pm$ 0.08 & 13.01 & 0.45 $\pm$ 0.01 & 1.95 $\pm$ 0.13 \\ 
274 & Gaia DR2 3462990284900676096 & 4860.85 $\pm$ 73.52 & 2.50 $\pm$ 0.18 & 0.03 $\pm$ 0.07 & 13.26 & 1.54 $\pm$ 0.08 & 1.94 $\pm$ 0.13 \\ 
275 & Gaia DR2 5915905270823073280 & 4783.33 $\pm$ 68.88 & 2.68 $\pm$ 0.18 & 0.08 $\pm$ 0.07 & 12.87 & 1.65 $\pm$ 0.07 & 1.94 $\pm$ 0.13 \\ 
276 & Gaia DR2 4634833353827302912 & 4670.48 $\pm$ 93.32 & 2.27 $\pm$ 0.18 &-0.36 $\pm$ 0.07 & 12.25 & 1.68 $\pm$ 0.03 & 1.94 $\pm$ 0.13 \\ 
277 & Gaia DR2 6750541665543575040 & 5097.21 $\pm$ 63.56 & 3.29 $\pm$ 0.18 & 0.13 $\pm$ 0.07 & 12.36 & 1.06 $\pm$ 0.03 & 1.94 $\pm$ 0.13 \\ 
278 & Gaia DR2 6462052526200514048 & 5187.33 $\pm$ 63.99 & 3.75 $\pm$ 0.18 & 0.15 $\pm$ 0.08 & 13.07 & 0.53 $\pm$ 0.02 & 1.94 $\pm$ 0.13 \\ 
279 & Gaia DR2 6726269083069826816 & 4765.87 $\pm$ 58.03 & 2.39 $\pm$ 0.16 & 0.02 $\pm$ 0.06 & 12.50 & 1.67 $\pm$ 0.09 & 1.93 $\pm$ 0.12 \\
280 & Gaia DR2 6289585296930463744 & 4620.54 $\pm$ 84.76 & 2.17 $\pm$ 0.18 &-0.35 $\pm$ 0.07 & 12.54 & 1.61 $\pm$ 0.05 & 1.93 $\pm$ 0.13 \\ 
281 & Gaia DR2 6107232942736558976 & 4691.86 $\pm$ 54.30 & 2.17 $\pm$ 0.15 &-0.37 $\pm$ 0.06 & 12.15 & 1.68 $\pm$ 0.07 & 1.93 $\pm$ 0.12 \\ 
282 & Gaia DR2 6892209198998772608 & 4762.65 $\pm$ 65.96 & 2.23 $\pm$ 0.17 &-0.22 $\pm$ 0.07 & 12.47 & 1.62 $\pm$ 0.07 & 1.92 $\pm$ 0.12 \\ 
283 & Gaia DR2 6086054935738957824 & 4698.35 $\pm$ 60.78 & 2.39 $\pm$ 0.17 &-0.08 $\pm$ 0.07 & 12.37 & 1.65 $\pm$ 0.07 & 1.92 $\pm$ 0.13 \\ 
284 & Gaia DR2 6199127547605022080 & 4769.17 $\pm$ 51.71 & 2.73 $\pm$ 0.15 & 0.28 $\pm$ 0.06 & 12.43 & 1.49 $\pm$ 0.12 & 1.92 $\pm$ 0.12 \\ 
285 & Gaia DR2 5914566409254818816 & 4994.99 $\pm$ 42.74 & 2.91 $\pm$ 0.13 &-0.12 $\pm$ 0.05 & 12.09 & 1.46 $\pm$ 0.04 & 1.92 $\pm$ 0.12 \\ 
286 & Gaia DR2 5228226516490660608 & 5003.83 $\pm$ 51.67 & 2.61 $\pm$ 0.16 & 0.03 $\pm$ 0.06 & 10.84 & 1.68 $\pm$ 0.02 & 1.92 $\pm$ 0.12 \\ 
287 & Gaia DR2 5635224523762161920 & 4743.68 $\pm$ 53.87 & 2.57 $\pm$ 0.16 & 0.12 $\pm$ 0.06 & 11.51 & 1.44 $\pm$ 0.03 & 1.92 $\pm$ 0.12 \\ 
288 & Gaia DR2 6791671715479344000 & 5004.36 $\pm$ 68.03 & 3.61 $\pm$ 0.19 &-0.20 $\pm$ 0.08 & 13.34 & 0.52 $\pm$ 0.02 & 1.92 $\pm$ 0.13 \\ 
289 & Gaia DR2 6102457488859232768 & 4762.85 $\pm$ 69.87 & 2.43 $\pm$ 0.19 & 0.11 $\pm$ 0.08 & 12.50 & 1.89 $\pm$ 0.12 & 1.91 $\pm$ 0.13 \\ 
290 & Gaia DR2 6674593209090865152 & 4797.03 $\pm$ 68.12 & 2.21 $\pm$ 0.19 &-0.42 $\pm$ 0.08 & 12.90 & 1.83 $\pm$ 0.12 & 1.90 $\pm$ 0.13 \\ 
291 & Gaia DR2 4345385887026498816 & 4957.41 $\pm$ 93.14 & 2.40 $\pm$ 0.18 &-0.27 $\pm$ 0.07 & 13.50 & 1.51 $\pm$ 0.06 & 1.90 $\pm$ 0.13 \\ 
292 & Gaia DR2 3459859459900563200 & 4992.39 $\pm$ 86.56 & 2.57 $\pm$ 0.19 &-0.61 $\pm$ 0.08 & 14.00 & 1.88 $\pm$ 0.12 & 1.89 $\pm$ 0.13 \\ 
293 & Gaia DR2 5389611183939763200 & 4715.61 $\pm$ 55.93 & 2.62 $\pm$ 0.16 & 0.16 $\pm$ 0.06 & 12.76 & 1.58 $\pm$ 0.05 & 1.89 $\pm$ 0.12 \\ 
294 & Gaia DR2 5370788335148341504 & 4722.85 $\pm$ 46.58 & 2.15 $\pm$ 0.14 & 0.14 $\pm$ 0.05 & 10.87 & 1.98 $\pm$ 0.06 & 1.89 $\pm$ 0.12 \\ 
295 & Gaia DR2 3237165607464496512 & 4833.19 $\pm$ 69.13 & 2.42 $\pm$ 0.17 &-0.07 $\pm$ 0.07 & 12.78 & 1.80 $\pm$ 0.08 & 1.88 $\pm$ 0.13 \\ 
296 & Gaia DR2 6008987627437263360 & 4846.39 $\pm$ 52.15 & 2.47 $\pm$ 0.16 & 0.02 $\pm$ 0.06 & 12.52 & 1.55 $\pm$ 0.05 & 1.88 $\pm$ 0.12 \\ 
297 & Gaia DR2 6009401971515883392 & 4836.57 $\pm$ 63.24 & 2.48 $\pm$ 0.17 &-0.28 $\pm$ 0.07 & 11.42 & 1.54 $\pm$ 0.06 & 1.88 $\pm$ 0.13 \\ 
298 & Gaia DR2 6112594711191576832 & 4810.96 $\pm$ 96.54 & 2.52 $\pm$ 0.20 &-0.21 $\pm$ 0.08 & 13.25 & 1.75 $\pm$ 0.12 & 1.87 $\pm$ 0.14 \\ 
299 & Gaia DR2 6672740424623847936 & 4640.50 $\pm$ 51.23 & 2.05 $\pm$ 0.15 &-0.52 $\pm$ 0.06 & 12.35 & 2.06 $\pm$ 0.11 & 1.87 $\pm$ 0.12 \\ 
300 & Gaia DR2 6095651580603667840 & 4649.68 $\pm$ 57.89 & 2.21 $\pm$ 0.17 &-0.28 $\pm$ 0.07 & 11.74 & 2.03 $\pm$ 0.07 & 1.87 $\pm$ 0.12 \\
\hline
\end{tabular}
\end{table*}

\begin{table*}
\begin{center}
\contcaption{} \end{center}
\begin{tabular}{cccccccccccccc}
\hline
\hline
S.No.&Object ID& Teff & log{\it g} & [Fe/H] & m$_v$ & L/L$_\odot$ & A(Li)\\
 & & (K)& (dex) & (dex)& (mag) & (dex) & (dex)\\
\hline
\hline
301 & Gaia DR2 5215362333516851584 & 4733.07 $\pm$ 85.14 & 2.24 $\pm$ 0.17 &-0.32 $\pm$ 0.07 & 12.17 & 1.62 $\pm$ 0.03 & 1.87 $\pm$ 0.13 \\ 
302 & Gaia DR2 5225119464122495104 & 4752.39 $\pm$ 55.27 & 2.39 $\pm$ 0.16 & 0.04 $\pm$ 0.06 & 12.58 & 1.47 $\pm$ 0.03 & 1.87 $\pm$ 0.12 \\ 
303 & Gaia DR2 5776561172933957888 & 4915.69 $\pm$ 59.25 & 3.14 $\pm$ 0.17 &-0.04 $\pm$ 0.07 & 12.77 & 1.18 $\pm$ 0.02 & 1.87 $\pm$ 0.13 \\ 
304 & Gaia DR2 6081354901485751168 & 5148.79 $\pm$ 91.21 & 3.63 $\pm$ 0.19 & 0.11 $\pm$ 0.08 & 13.85 & 0.66 $\pm$ 0.03 & 1.87 $\pm$ 0.13 \\ 
305 & Gaia DR2 5369251733295804032 & 5196.19 $\pm$ 62.21 & 3.89 $\pm$ 0.18 & 0.24 $\pm$ 0.07 & 13.36 & 0.45 $\pm$ 0.01 & 1.87 $\pm$ 0.13 \\ 
306 & Gaia DR2 5634914908156544896 & 4859.02 $\pm$ 64.72 & 2.61 $\pm$ 0.18 & 0.07 $\pm$ 0.07 & 13.40 & 1.75 $\pm$ 0.05 & 1.86 $\pm$ 0.13 \\ 
307 & Gaia DR2 5370273385753624448 & 4684.29 $\pm$ 76.94 & 2.47 $\pm$ 0.20 & 0.07 $\pm$ 0.08 & 11.97 & 1.59 $\pm$ 0.04 & 1.86 $\pm$ 0.14 \\ 
308 & Gaia DR2 5417103819359766656 & 4864.13 $\pm$ 60.47 & 2.42 $\pm$ 0.17 &-0.12 $\pm$ 0.07 & 11.81 & 1.60 $\pm$ 0.03 & 1.86 $\pm$ 0.13 \\ 
309 & Gaia DR2 578047695475401472  & 5198.70 $\pm$ 75.81 & 3.78 $\pm$ 0.20 & 0.37 $\pm$ 0.08 & 13.42 & 0.61 $\pm$ 0.03 & 1.86 $\pm$ 0.14 \\ 
310 & Gaia DR2 6461096844437325184 & 5182.80 $\pm$ 59.47 & 3.89 $\pm$ 0.17 & 0.37 $\pm$ 0.07 & 12.73 & 0.50 $\pm$ 0.02 & 1.86 $\pm$ 0.13 \\ 
311 & Gaia DR2 6735003710852578304 & 4930.09 $\pm$ 23.19 & 2.41 $\pm$ 0.08 &-0.07 $\pm$ 0.03 &  9.26 & 1.91 $\pm$ 0.03 & 1.86 $\pm$ 0.11 \\ 
312 & Gaia DR2 3287566945603422848 & 4958.70 $\pm$ 56.60 & 2.28 $\pm$ 0.16 &-0.57 $\pm$ 0.06 & 12.22 & 1.63 $\pm$ 0.06 & 1.85 $\pm$ 0.12 \\ 
313 & Gaia DR2 5908145815408175232 & 4768.10 $\pm$ 69.77 & 2.79 $\pm$ 0.19 & 0.16 $\pm$ 0.08 & 13.81 & 1.54 $\pm$ 0.06 & 1.84 $\pm$ 0.13 \\ 
314 & Gaia DR2 5243639676641783808 & 4986.67 $\pm$ 53.12 & 2.50 $\pm$ 0.16 &-0.07 $\pm$ 0.06 & 10.85 & 1.76 $\pm$ 0.04 & 1.84 $\pm$ 0.12 \\ 
315 & Gaia DR2 5782285367906704512 & 4950.70 $\pm$ 52.05 & 2.61 $\pm$ 0.16 & 0.05 $\pm$ 0.06 & 10.73 & 1.69 $\pm$ 0.02 & 1.84 $\pm$ 0.12 \\ 
316 & Gaia DR2 6396539225371784704 & 5140.81 $\pm$ 57.40 & 3.90 $\pm$ 0.17 & 0.31 $\pm$ 0.07 & 13.06 & 0.40 $\pm$ 0.01 & 1.84 $\pm$ 0.13 \\ 
317 & Gaia DR2 4238064412465062912 & 5105.94 $\pm$ 60.57 & 3.57 $\pm$ 0.15 & 0.19 $\pm$ 0.06 & 10.79 & 0.75 $\pm$ 0.02 & 1.84 $\pm$ 0.12 \\ 
318 & Gaia DR2 5445306464409127424 & 4730.57 $\pm$ 53.13 & 2.80 $\pm$ 0.16 & 0.34 $\pm$ 0.06 & 13.18 & 1.52 $\pm$ 0.06 & 1.83 $\pm$ 0.12 \\ 
319 & Gaia DR2 6069512787599338112 & 4888.85 $\pm$ 64.78 & 2.37 $\pm$ 0.18 &-0.08 $\pm$ 0.07 & 10.99 & 1.58 $\pm$ 0.05 & 1.83 $\pm$ 0.13 \\ 
320 & Gaia DR2 5590608811508112384 & 4917.13 $\pm$ 62.39 & 2.54 $\pm$ 0.18 &-0.04 $\pm$ 0.07 & 12.80 & 1.63 $\pm$ 0.05 & 1.82 $\pm$ 0.13 \\ 
321 & Gaia DR2 6135018975959484544 & 4853.30 $\pm$ 58.58 & 2.48 $\pm$ 0.17 & 0.03 $\pm$ 0.07 & 12.67 & 1.52 $\pm$ 0.07 & 1.82 $\pm$ 0.12 \\ 
322 & Gaia DR2 4297442541500760320 & 4819.18 $\pm$ 50.32 & 2.55 $\pm$ 0.15 & 0.04 $\pm$ 0.06 & 11.17 & 1.74 $\pm$ 0.06 & 1.82 $\pm$ 0.12 \\ 
323 & Gaia DR2 6105298077149947776 & 5193.09 $\pm$ 68.19 & 3.89 $\pm$ 0.19 & 0.32 $\pm$ 0.08 & 13.91 & 0.47 $\pm$ 0.03 & 1.82 $\pm$ 0.13 \\ 
324 & Gaia DR2 2903941845411604224 & 5152.86 $\pm$ 61.57 & 3.91 $\pm$ 0.17 & 0.17 $\pm$ 0.07 & 13.19 & 0.54 $\pm$ 0.02 & 1.82 $\pm$ 0.13 \\ 
325 & Gaia DR2 5416873987069851008 & 4933.56 $\pm$ 62.48 & 3.42 $\pm$ 0.18 & 0.33 $\pm$ 0.07 & 11.55 & 0.88 $\pm$ 0.01 & 1.82 $\pm$ 0.13 \\ 
326 & Gaia DR2 6123223449572807424 & 4759.80 $\pm$ 68.24 & 2.50 $\pm$ 0.17 & 0.08 $\pm$ 0.07 & 13.26 & 1.65 $\pm$ 0.10 & 1.81 $\pm$ 0.12 \\ 
327 & Gaia DR2 3137948018934622464 & 4858.64 $\pm$ 65.34 & 2.45 $\pm$ 0.16 &-0.20 $\pm$ 0.06 & 10.99 & 1.83 $\pm$ 0.05 & 1.81 $\pm$ 0.12 \\ 
328 & Gaia DR2 6082171735546777856 & 4830.19 $\pm$ 54.79 & 2.73 $\pm$ 0.16 & 0.18 $\pm$ 0.06 & 11.24 & 1.61 $\pm$ 0.07 & 1.81 $\pm$ 0.12 \\ 
329 & Gaia DR2 6385111412612497920 & 5191.38 $\pm$ 64.57 & 4.01 $\pm$ 0.16 & 0.34 $\pm$ 0.06 & 12.39 & 0.48 $\pm$ 0.01 & 1.81 $\pm$ 0.12 \\ 
330 & Gaia DR2 6688625417003949056 & 5196.40 $\pm$ 49.39 & 3.71 $\pm$ 0.15 &-0.02 $\pm$ 0.06 & 12.09 & 0.51 $\pm$ 0.02 & 1.81 $\pm$ 0.12 \\ 
331 & Gaia DR2 6037492191509529600 & 4195.12 $\pm$ 74.10 & 1.95 $\pm$ 0.19 &-0.24 $\pm$ 0.08 & 13.90 & 2.08 $\pm$ 0.11 & 1.80 $\pm$ 0.13 \\ 
332 & Gaia DR2 6396441643714574208 & 4512.14 $\pm$ 93.63 & 2.00 $\pm$ 0.20 &-0.20 $\pm$ 0.08 & 12.94 & 2.07 $\pm$ 0.08 & 1.80 $\pm$ 0.14 \\ 
333 & Gaia DR2 4334137470757431680 & 4860.61 $\pm$ 81.18 & 2.79 $\pm$ 0.20 &-0.04 $\pm$ 0.08 & 14.00 & 1.12 $\pm$ 0.34 & 1.80 $\pm$ 0.13 \\ 
334 & Gaia DR2 6852673513940765952 & 5022.58 $\pm$ 73.46 & 3.76 $\pm$ 0.20 & 0.02 $\pm$ 0.08 & 13.57 & 0.35 $\pm$ 0.02 & 1.80 $\pm$ 0.14 \\ 
335 & Gaia DR2 3461057652697403776 & 5190.11 $\pm$ 62.03 & 3.85 $\pm$ 0.18 & 0.38 $\pm$ 0.07 & 12.68 & 0.51 $\pm$ 0.04 & 1.80 $\pm$ 0.13 \\ 
\hline
\end{tabular}
\end{table*}

\bsp	
\label{lastpage}
\end{document}